\documentclass[american,pre,onecolumn]{revtex4}
\usepackage[T1]{fontenc}
\usepackage[latin9]{inputenc}
\usepackage{float}
\usepackage{amsmath}
\usepackage{amssymb}
\usepackage{graphicx}

\makeatletter

\providecommand{\tabularnewline}{\\}

\@ifundefined{textcolor}{}
{%
 \definecolor{BLACK}{gray}{0}
 \definecolor{WHITE}{gray}{1}
 \definecolor{RED}{rgb}{1,0,0}
 \definecolor{GREEN}{rgb}{0,1,0}
 \definecolor{BLUE}{rgb}{0,0,1}
 \definecolor{CYAN}{cmyk}{1,0,0,0}
 \definecolor{MAGENTA}{cmyk}{0,1,0,0}
 \definecolor{YELLOW}{cmyk}{0,0,1,0}
}

\usepackage{graphicx}
\graphicspath{ {figs/} {../figs/} }

\makeatother

\usepackage{babel}
\begin{document}
\title{The fourth virial coefficient for hard spheres in even dimension}
\author{Ignacio Urrutia}
\email{iurrutia@cnea.gov.ar}

\affiliation{Departamento de Física de la Materia Condensada, Centro Atómico Constituyentes,
CNEA, Av.Gral.~Paz 1499, 1650 Pcia.~de Buenos Aires, Argentina}
\affiliation{Instituto de Nanociencia y Nanotecnología, CONICET-CNEA, CAC.}
\begin{abstract}
The fourth virial coefficient is calculated exactly for a fluid of
hard spheres in even dimensions. For this purpose the complete star
cluster integral is expressed as the sum of two three-folded integrals
only involving spherical angular coordinates. These integrals are
solved analytically for any even dimension $d$, and working with
existing expressions for the other terms of the fourth cluster integral,
we obtain an expression for the fourth virial coefficient $B_{4}(d)$
for even $d$. It reduces to the sum of a finite number of simple
terms that increases with $d$.
\end{abstract}
\maketitle

\section{Introduction}

An essential problem in statistical mechanics of fluids at equilibrium
is how to obtain its equation of state (EOS), given the interaction
potential of the particles system. This basic question has not an
answer yet. However, for low density systems, one has the pressure
virial series
\begin{equation}
P/k_{B}T=\rho+\sum_{n=2}^{\infty}B_{n}\rho^{n}\:,\label{eq:VirialSeries}
\end{equation}
the series expansion of the EOS in powers of the density ($\rho$
is the density, $T$ temperature and $k_{B}$ the Boltzmann constant).
Here, $B_{n}$ is the $n$-th coefficient of the series studied by
Mayer and others, an integral over the position of $n$ particles.
In such coefficients, the integrand is interpreted as the double-connected
or irreducible cluster where particles are connected by the Mayer
$f$ function 
\begin{equation}
f_{ij}=f(r_{ij})=\exp\left[-W(r_{ij})/k_{B}T\right]-1\:,\label{eq:fMayer}
\end{equation}
and $W$ is the pair interaction potential between particles. The
lowest order virial coefficients are
\begin{equation}
B_{2}=-\frac{1}{2V}\int f_{12}d\mathbf{r}_{1}d\mathbf{r}_{2}=-\frac{1}{2}\:\includegraphics[width=0.5cm]{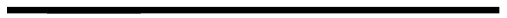}\:,\label{eq:B2diagram}
\end{equation}
\begin{equation}
B_{3}=-\frac{1}{3V}\int f_{12}f_{13}f_{23}d\mathbf{r}_{1}d\mathbf{r}_{2}d\mathbf{r}_{3}=-\frac{1}{3}\:\includegraphics[width=0.5cm]{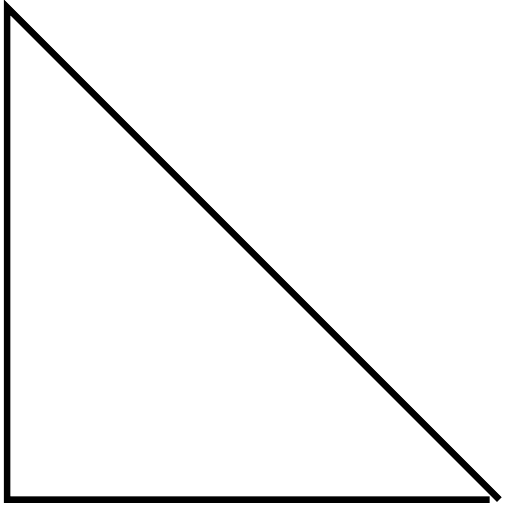}\:\label{eq:B3diagram}
\end{equation}
and the Mayer graph representation of the fourth cluster integral
is 
\begin{equation}
B_{4}=-\frac{1}{8}\includegraphics[width=0.5cm]{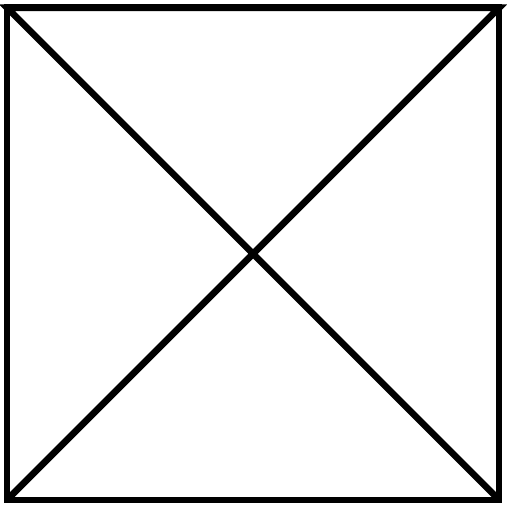}-\frac{3}{4}\:\includegraphics[width=0.5cm]{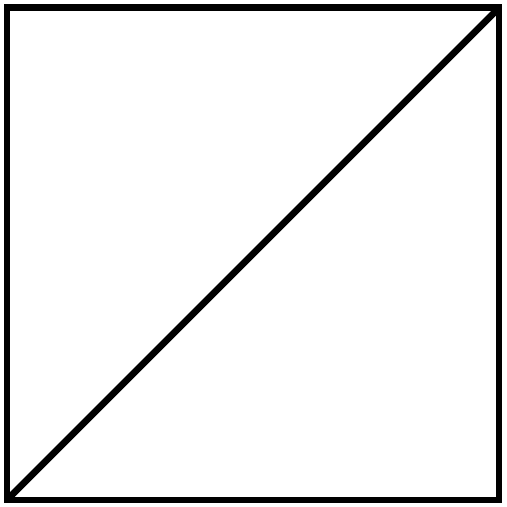}-\frac{3}{8}\:\includegraphics[width=0.5cm]{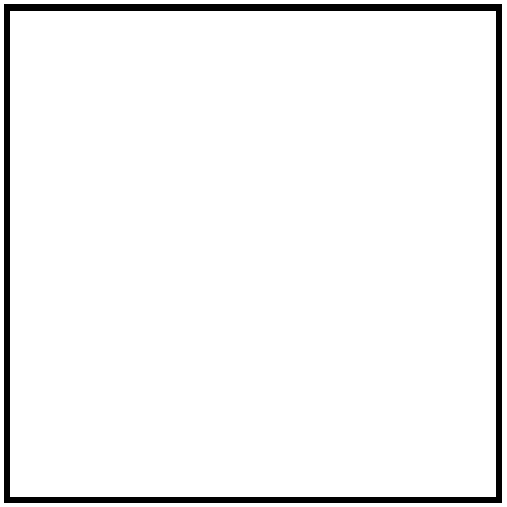}\:.\label{eq:B4MayerDiag}
\end{equation}

Hard sphere (HS) system is a minimal model of particle-particle interaction
that introduces the excluded volume expressing that the minimum distance
between two particles is finite. For HS the two-body interaction potential
is $W(r_{ij})=0$ if $r_{ij}>\sigma$ and $W(r_{ij})=+\infty$ if
$r_{ij}<\sigma$, with $\sigma$ the HS diameter. The origin of HS
as a model of fluid could probably be traced back nearly 150 years
ago from van der Waals theory.\citep{vanderWaals_1873} The question
of found the \emph{exact} EOS of the HS, the minimal model of fluid,
remains up today as an open problem. The hard sphere fluid was studied
not only in dimensions three and two, but also in arbitrary integer
dimension $d$. The virial expansion of the HS EOS has been studied
for more than a century, but has not been solved. The more recent
advances were done fifteen years ago. Clisby and McCoy obtained $B_{4}$
for even dimensions $d=4,6,8,10,12$ in 2004 and Lyberg calculated
$B_{4}$ for odd dimensions $d=5,7,9,11$ in 2005.\citep{Clisby_2004,Lyberg_2005}
Previously, $B_{4}$ for dimension 3 was successfully calculated by
Boltzmann\citep{Boltzmann_1899} and van Laar in 1899.\citep{vanLaar_1899_b}
(See \citep{Nairn_1972,Kilpatrick_1971,Santos2016} for the interesting
history of the calculation of $B_{4}$ in 1899, that includes the
contribution of van der Waals. The complete reference to the original
papers is given in Ref. \citep{Luban_1982}.) For dimension 2 it was
calculated independently by Rowlinson\citep{Rowlinson_1964_b} and
Hemmer\citep{Hemmer_1965} in 1964. In the present work we solve $B_{4}$
for any even dimension.

Low order virial coefficients of HS are well known, $B_{2}(d)$ is
half the volume of a sphere in the $d$ dimensional space,
\begin{equation}
B_{2}=\sigma^{d}\frac{\pi^{d/2}}{2\Gamma(d/2+1)}\:.\label{eq:B2-hs}
\end{equation}
$B_{3}$ was calculated initially for $d=3$ and $d=2$,\citep{Boltzmann_1899,Tonks_1936}
and latter for arbitrary dimension. It reduces to
\begin{equation}
\frac{B_{3}}{B_{2}^{2}}=\frac{2\,\sigma^{2d}}{B\left(\frac{d+1}{2},\frac{1}{2}\right)}B_{3/4}\left(\frac{d+1}{2},\frac{1}{2}\right)\:,\label{eq:B3-hs}
\end{equation}
where $B\left(a,b\right)=\frac{\Gamma\left(a\right)\Gamma\left(b\right)}{\Gamma\left(a+b\right)}$
is the beta function and $B_{x}\left(a,b\right)$ is the incomplete
beta function, in particular $B_{3/4}\left(\frac{d+1}{2},\frac{1}{2}\right)=2\int_{0}^{\pi/3}\left(\sin\varphi\right)^{d}d\varphi$.
Exact expressions for two of the three diagrams of $B_{4}$ are known,
the Mayer cluster integrals \includegraphics[viewport=0bp 10bp 146bp 146bp,width=0.5cm]{graph-b4-sq}
and \includegraphics[viewport=0bp 0bp 146bp 146bp,width=0.5cm]{graph-b4-sqd}.
There exist several expressions for both, Luban and Baram calculated
them for arbitrary dimension,\citep{Luban_1982} and found
\[
\frac{\includegraphics[width=0.5cm]{graph-b4-sq.pdf}}{B_{2}^{3}}=\frac{2^{d+6}d}{(d+1)^{2}}\frac{B\left(d,\frac{d}{2}+1\right)}{B\left(\frac{d+1}{2},\frac{1}{2}\right)^{2}}\,_{3}F_{2}\left(\frac{1}{2},1,\frac{1-d}{2};\frac{d+3}{2},\frac{d+3}{2};1\right)\:,
\]
where $_{3}F_{2}$ is the generalized hypergeometric function. Based
on the expression obtained by Luban and later modified by Joslin \citep{Luban_1982,Joslin_1982}
we found
\begin{equation}
\frac{\includegraphics[width=0.5cm]{graph-b4-sqd.pdf}}{B_{2}^{3}}=-\frac{2^{d+3}d}{\left[B\left(\frac{d+1}{2},\frac{1}{2}\right)\right]^{2}}\intop_{0}^{1/2}x^{d-1}\left[B_{1-x^{2}}\left(\frac{d+1}{2},\frac{1}{2}\right)\right]^{2}\,dx\:,\label{eq:B4Sqrd}
\end{equation}
with $B_{1-x^{2}}\left(\frac{d+1}{2},\frac{1}{2}\right)=2\int_{0}^{\arccos x}\left(\sin\varphi\right)^{d}d\varphi$.
On the opposite, no general expression is known for the complete star
cluster integral \includegraphics[viewport=0bp 0bp 146bp 146bp,width=0.5cm]{graph-b4-fs}.

In the case of even dimension $d=2\nu$ (being $\nu$ a positive integer)
one have the following closed expressions for $B_{2}$ and $B_{3}$
\begin{eqnarray}
B_{2}=\frac{\sigma^{2\nu}\pi^{\nu}}{2\,\nu!} & \;\;\textrm{and}\;\; & \frac{B_{3}}{B_{2}^{2}}=\frac{4}{3}-\frac{\sqrt{3}}{\pi}\sum_{k=1}^{\nu}\frac{(k-1)!}{(2k-1)\text{!!}}\left(\frac{3}{2}\right)^{k-1}\:.\label{eq:B2yB3Even}
\end{eqnarray}
To evaluate $B_{4}$ for arbitrary large even dimensions $d=2\nu$
it is convenient to give a closed form expression for each star Mayer
diagrams of four points. Joslin simplified an expression found by
Luban and Baram to obtain\citep{Joslin_1982}

\begin{equation}
\frac{\includegraphics[width=0.5cm]{graph-b4-sq.pdf}}{B_{2}^{3}}=8-\frac{8}{\pi^{2}}\sum_{k=1}^{\nu}\frac{(k-1)!2^{k-1}}{k(2k-1)!!}\left[4+\frac{\nu\,(2\nu)!!\,4^{k}k!\,(\nu+k-1)!}{(2\nu-1)!!\,(\nu+2k)!}\right]\:.\label{eq:B4SqrE}
\end{equation}
To derive a closed form expression of \includegraphics[width=0.5cm]{graph-b4-sqd}
for even dimension we start from Eq. \eqref{eq:B4Sqrd}. In the Appendix
\ref{Appsec:B4sqrdsh} we solve this integral using change of variables
and several identities for the integral of powers of trigonometric
functions taken from Ref. \citep{GradshteynRyzhik2007}. Here we
sate the result
\begin{equation}
\frac{\includegraphics[width=0.5cm]{graph-b4-sqd.pdf}}{B_{2}^{3}}=-8\biggl(1-\frac{\sqrt{3}}{2\pi}R_{0}\biggr)^{2}+\frac{4}{\pi^{2}}\left[R_{1}+\frac{8(2\nu)!!}{(2\nu-1)!!}R_{2}\right]\:,\label{eq:B4SqrdE}
\end{equation}
with
\begin{eqnarray}
R_{0} & = & \sum_{k=1}^{\nu}\frac{(k-1)!}{(2k-1)!!}\left(\frac{3}{2}\right)^{k-1}\;,\nonumber \\
R_{1} & = & \sum_{k=1}^{\nu}\frac{(k-1)!}{k(2k-1)!!}\left(\frac{3}{2}\right)^{k}\;,\label{eq:R123}\\
R_{2} & = & \nu!\sum_{k=1}^{\nu}\frac{(k-1)!2^{k-1}}{(2k-1)!!(2\nu+k)!}\biggl[4^{\nu}(\nu+k-1)!-\frac{3^{k+\nu}}{4^{k}}\sum_{j=0}^{\nu}\frac{(\nu+k+j-1)!}{4^{j}j!}\biggr]\;.\nonumber 
\end{eqnarray}
Here, $R_{0}$, $R_{1}$ and $R_{2}$ are rational numbers. The Eq.
\eqref{eq:B4SqrdE} is an explicit form that, for a given even dimension,
only involves a finite number of simple operations (sum, product,
quotient).

The rest of this paper is devoted to transform the complete star integral
to a triple integral for any dimension $d$ and to solve the complete
star integral to obtain $B_{4}$ for even dimension. It is organized
as follows. In Sec. \ref{sec:Transf-FullStarDiagram} we transform
the complete star diagram and find a new expression for it in terms
of two three-folded integrals. In Sec. \ref{sec:The-three-foldedIntegral}
we solve each of these integrals for even dimension $d$ and reach
an explicit form of the complete star in terms of a finite sum. Some
mathematical details of the calculations are relegated to the appendices.
In Sec. \ref{sec:Results} we give the exact expression of $B_{4}$
and evaluate it for some even dimensions as large as $d=1000$. Sec.
\ref{sec:Final-remarks} is devoted to some final remarks.

\section{The transformation of the complete star\label{sec:Transf-FullStarDiagram}}

The full star diagram of four vertex contains six $f$-bonds. These
bonds connect each pair of vertex. For HS it takes the form $f_{ij}=-\Theta\left(\sigma-r_{ij}\right)$,
being $\Theta\left(x\right)$ the Heaviside function, with $\Theta\left(x\right)=1$
if $x\geq0$ and $\Theta\left(x\right)=0$ if $x<0$. The symmetric
form of the star integral is
\begin{equation}
\includegraphics[width=0.5cm]{graph-b4-fs.pdf}=V^{-1}\iiiint f_{12}f_{13}f_{14}f_{23}f_{24}f_{34}d\mathbf{r}_{1234}\:,\label{eq:B4s-fs}
\end{equation}
with $d\mathbf{r}_{1234}=d\mathbf{r}_{1}d\mathbf{r}_{2}d\mathbf{r}_{3}d\mathbf{r}_{4}$.
Here, each vertex represents a particle whose position is integrated
over the complete $d$-dimensional euclidean space, while each line
is a $f$-bond. The spatial translation of the rigid cluster produces
a volume term that can be integrated out (also, the rigid rotations
produce solid angle terms). Hence, 
\begin{eqnarray}
\includegraphics[width=0.5cm]{graph-b4-fs.pdf} & = & \iiiint f_{12}f_{13}f_{14}f_{23}f_{24}f_{34}d\mathbf{r}_{12}d\mathbf{r}_{13}d\mathbf{r}_{14}\label{eq:B4s-ypct}\\
 & = & \sigma^{3d}\mathrm{B_{4}^{s}}\:,\label{eq:B4s-sgm}
\end{eqnarray}
where $\mathrm{B_{4}^{s}}$ is a number. Our purpose is to find the
dependence of this number with $d$. Eq. \eqref{eq:B4s-ypct} could
also be rewritten in different forms by adopting a different set of
integrating variables, for example, by replacing $d\mathbf{r}_{12}d\mathbf{r}_{13}d\mathbf{r}_{14}$
with $d\mathbf{r}_{12}d\mathbf{r}_{23}d\mathbf{r}_{34}$.

In the following we present a reformulation of the full star diagram
that reduces it to three-folded integrals. On one hand, we take the
derivative of Eq. \eqref{eq:B4s-sgm} to obtain
\begin{equation}
\frac{\partial}{\partial\sigma}\:\includegraphics[width=0.5cm]{graph-b4-fs.pdf}=\frac{3d}{\sigma}\:\includegraphics[width=0.5cm]{graph-b4-fs.pdf}\:.\label{eq:B4s-Der0}
\end{equation}
On the other hand, through applying the derivative to Eq. \eqref{eq:B4s-fs}
in successive steps we transform \includegraphics[viewport=0bp 0bp 146bp 146bp,clip,width=0.5cm]{graph-b4-fs}
by noting that $\frac{\partial}{\partial\sigma}f_{ij}=-\delta\left(\sigma-r_{ij}\right)=-\delta_{ij}$
(with $\delta\left(x\right)$ the Dirac delta function) which introduce
the $\delta$-bonds (corresponding to $\frac{\partial}{\partial r}f_{ij}=f'_{ij}=\delta_{ij}$).
This is the rule to transform a $f$-bond to $\delta$-bond. This
process ends with the complete star diagram written as a sum over
all the $\delta$-simple-connected diagrams, where the non-$\delta$-bonds
remain unmodified and each diagram includes a weight function. To
draw the graphs we introduce the open-box link for the $\delta$-bond,
when it appears formally integrated, and the bold line for the same
bond once the coordinate $r_{ij}$ is integrated on. The transformation
applied to \includegraphics[viewport=0bp 0bp 146bp 146bp,clip,width=0.5cm]{graph-b4-fs}
results in two integrals each of them three folded.

We take the derivative of Eq. \eqref{eq:B4s-fs} giving $\frac{\partial}{\partial\sigma}\:\includegraphics[width=0.5cm]{graph-b4-fs.pdf}=-\frac{6}{V}\iiiint\delta_{12}f_{13}f_{14}f_{23}f_{24}f_{34}d\mathbf{r}_{1234}$.
Once we integrate on the $r_{12}$ degree of freedom, that fixes $r_{12}=\sigma$
and produces the factor $\sigma^{d-1}$, we obtain 
\begin{eqnarray}
\frac{\partial}{\partial\sigma}\:\includegraphics[width=0.5cm]{graph-b4-fs.pdf} & = & -6\:\includegraphics[width=0.5cm]{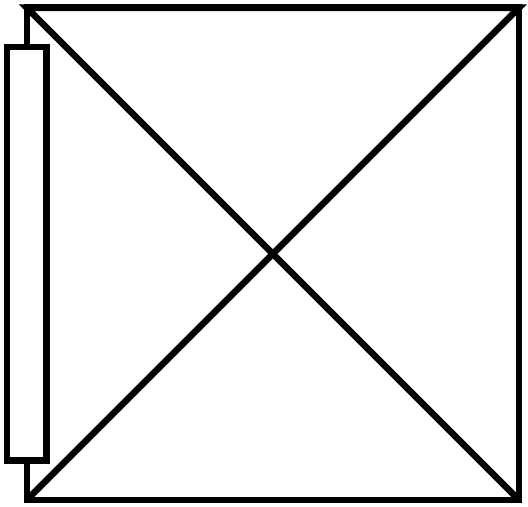}\:,\label{eq:B4s-Der1a}\\
 & = & -6\sigma^{d-1}{}_{*}\includegraphics[width=0.5cm]{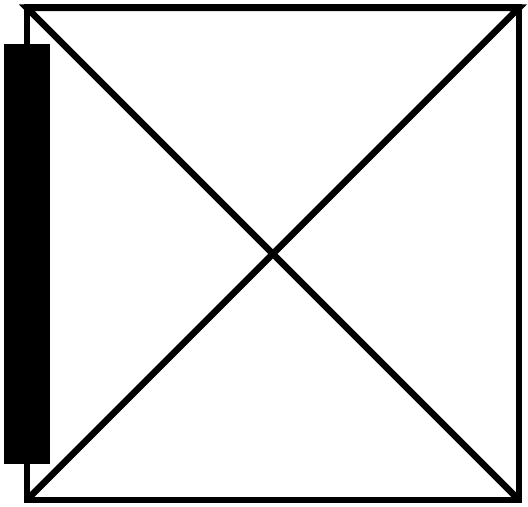}\:.\label{eq:B4s-Der1b}
\end{eqnarray}
Here, the star marks one of the vertex at the ends of the bold-bond
(any of them), that was a fixed position allowing the integration.
It is interesting to note that the integrated bond in \includegraphics[clip,width=0.5cm]{graph-b4-sgmI}
has not dependence on $\sigma$. We obtain 
\begin{equation}
\includegraphics[width=0.5cm]{graph-b4-fs.pdf}=-\frac{2\sigma^{d}}{d}\:{}_{*}\includegraphics[width=0.5cm]{graph-b4-sgmI.pdf}\:.\label{eq:B4s-resD1}
\end{equation}
The expression at the right of Eq. \eqref{eq:B4s-resD1}, once the
full rotational symmetry is integrated on (producing the solid angle
term), is equivalent to the two center formulation of the full star.\citep{deBoer_1949,Nijboer_1952}

(Second) We take the derivative $\frac{\partial}{\partial\sigma}$
at both sides of Eq. \eqref{eq:B4s-resD1}, which gives
\begin{equation}
\frac{3d}{\sigma}\:\includegraphics[width=0.5cm]{graph-b4-fs.pdf}=\frac{d}{\sigma}\:\includegraphics[width=0.5cm]{graph-b4-fs.pdf}-\frac{2\sigma^{d}}{d}\frac{\partial}{\partial\sigma}\:{}_{*}\includegraphics[width=0.5cm]{graph-b4-sgmI.pdf}\quad.\label{eq:B4-2a}
\end{equation}
   The last term is
\begin{eqnarray}
\frac{\partial}{\partial\sigma}\:{}_{*}\includegraphics[width=0.5cm]{graph-b4-sgmI.pdf} & = & -2\:{}_{*}\includegraphics[width=0.5cm]{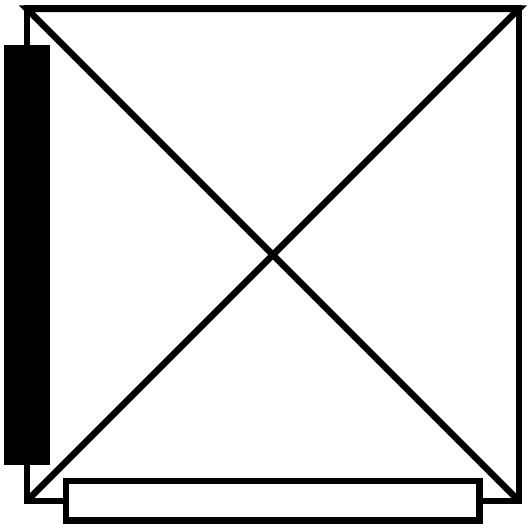}-2\:{}_{*}\includegraphics[width=0.5cm]{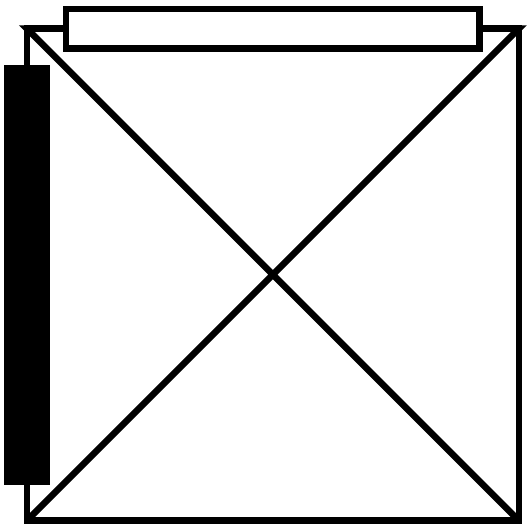}-{}_{*}\includegraphics[width=0.5cm]{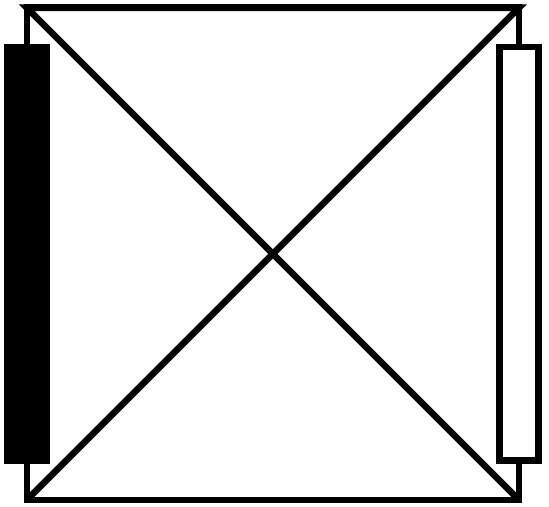}\nonumber \\
 & = & -\sigma^{d-1}\left(2\:{}_{**}\includegraphics[width=0.5cm]{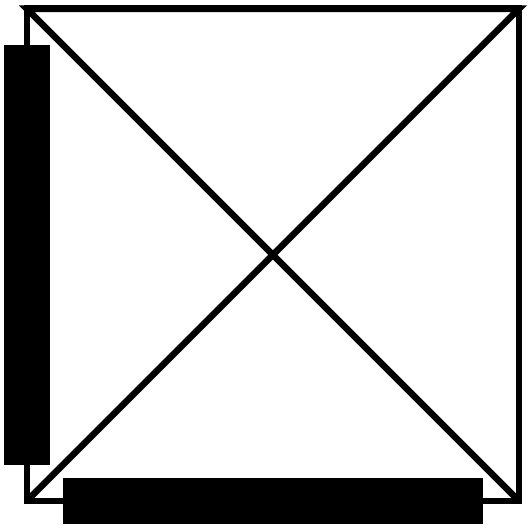}\left\langle 2-\cos\varphi\right\rangle +{}_{*}\includegraphics[width=0.5cm]{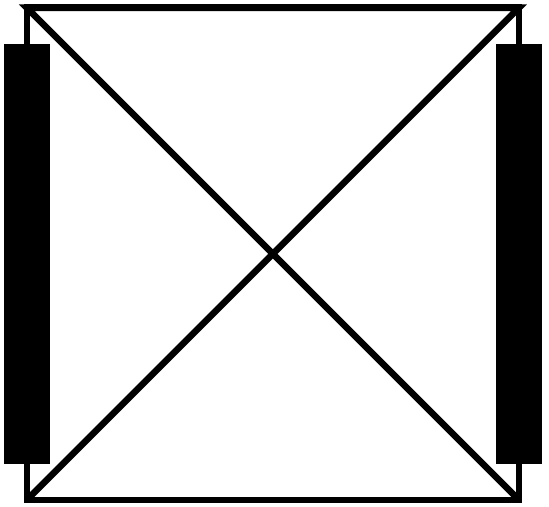}{}_{*}\right)\:,\label{eq:B22222}
\end{eqnarray}
where we used the identity
\begin{equation}
\includegraphics[height=0.5cm]{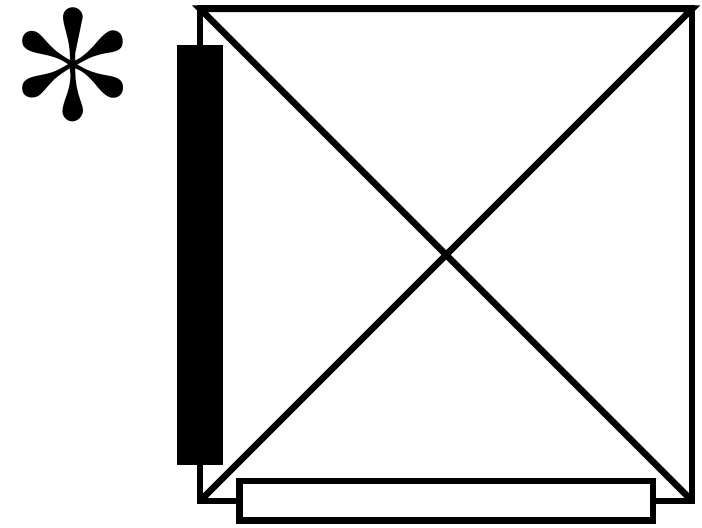}={}_{*}\includegraphics[width=0.5cm]{graph-b4-sgm1I1c.pdf}\left\langle 1-\cos\varphi\right\rangle \:.\label{eq:cosIdent}
\end{equation}
This relation is derived in the Appendix \ref{Appsec:CosTerm}. In
Eqs. \eqref{eq:B22222} and \eqref{eq:cosIdent} $\varphi$ corresponds
to the angle between both $\delta$-bonds. Note that term $\left\langle 2-\cos\varphi\right\rangle $
is in our notation a term $\left(2-\cos\varphi\right)$ inside the
diagram integrand (none dependence on the angular variables can appear
outside of the integral given that all angular degrees of freedom
are integrated). This notation using $\left\langle ...\right\rangle $
will be used from here on.

Replacing Eq. \eqref{eq:B22222} in Eq. \eqref{eq:B4-2a} we obtain
\begin{equation}
\includegraphics[width=0.5cm]{graph-b4-fs.pdf}=\frac{\sigma^{2d}}{d^{2}}\left(2\:{}_{*}\includegraphics[width=0.5cm]{graph-b4-sgm2I.pdf}\left\langle 2-\cos\varphi\right\rangle +{}_{*}\includegraphics[width=0.5cm]{graph-b4-sgm2aI.pdf}{}_{*}\right)\:.\label{eq:B4s-resD2}
\end{equation}

(Third) We take the derivative $\frac{\partial}{\partial\sigma}$
at both sides of Eq. \eqref{eq:B4s-resD2} which gives
\begin{equation}
\frac{3d}{\sigma}\:\includegraphics[width=0.5cm]{graph-b4-fs.pdf}=\frac{2d}{\sigma}\:\includegraphics[width=0.5cm]{graph-b4-fs.pdf}+\frac{\sigma^{2d}}{d^{2}}\frac{\partial}{\partial\sigma}\left(2\:{}_{*}\includegraphics[width=0.5cm]{graph-b4-sgm2I.pdf}\:\left\langle 2-\cos\varphi\right\rangle +{}_{*}\includegraphics[width=0.5cm]{graph-b4-sgm2aI.pdf}{}_{*}\right)\:.\label{eq:B4-3a}
\end{equation}
The derivative term in Eq. \eqref{eq:B4-3a} is

\begin{eqnarray*}
\frac{\partial}{\partial\sigma}\left(...\right) & = & -2\,{}_{*}\includegraphics[width=0.5cm]{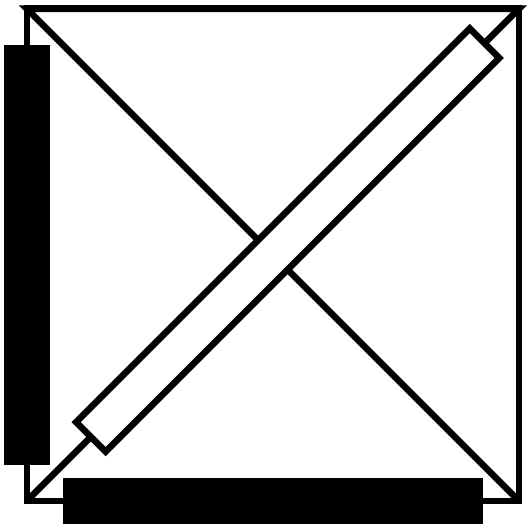}\left\langle 2-\cos\varphi\right\rangle -4\,{}_{*}\includegraphics[width=0.5cm]{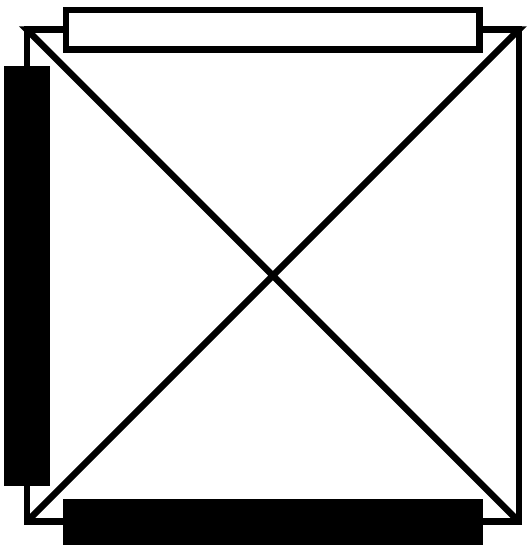}\left\langle 2-\cos\varphi\right\rangle -{}_{*}\includegraphics[width=0.5cm]{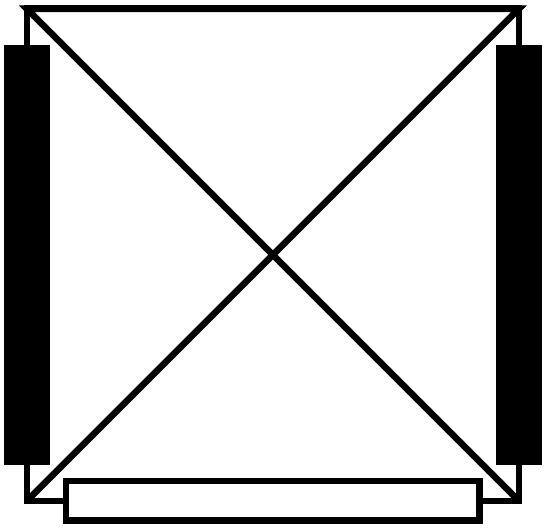}_{*}-{}_{*}\includegraphics[width=0.5cm]{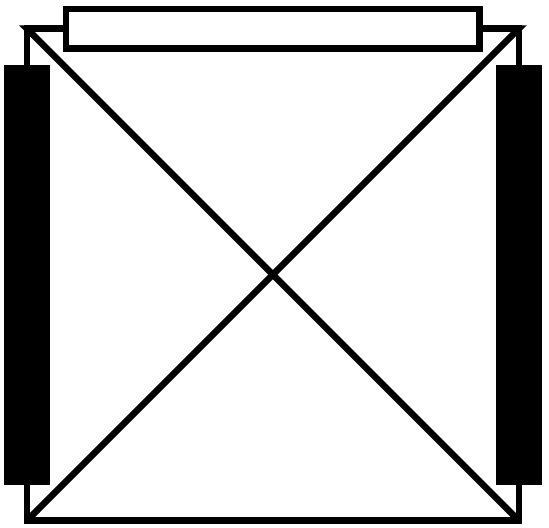}_{*}-2\,{}_{*}\includegraphics[width=0.5cm]{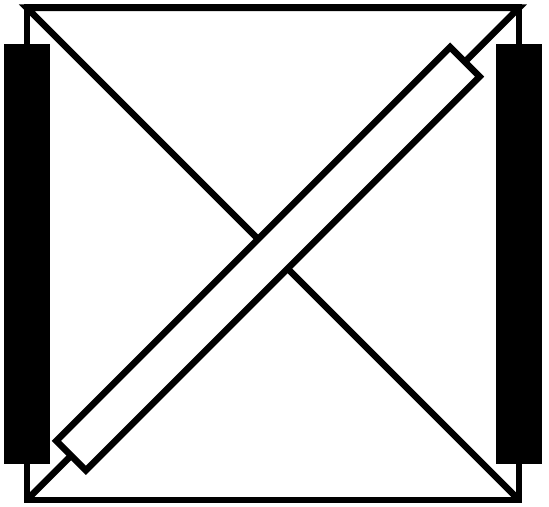}_{*}\:,\\
 & = & -\sigma^{d-1}\Bigl[2\,{}_{*}\includegraphics[width=0.5cm]{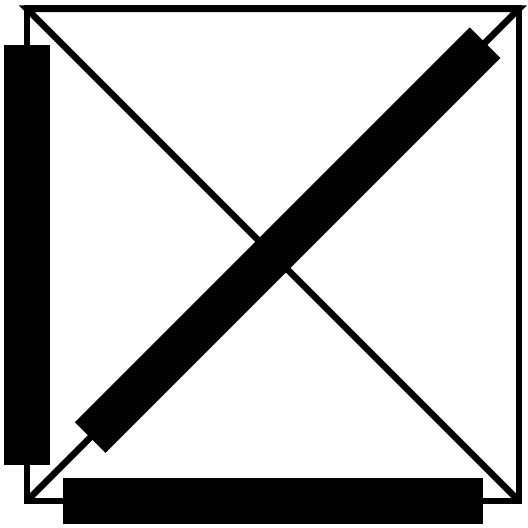}\left\langle 2-\cos\varphi\right\rangle +4\,{}_{*}\includegraphics[width=0.5cm]{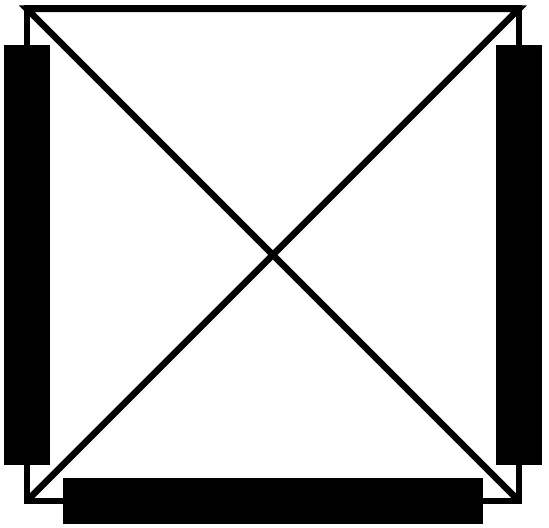}_{*}\,\left\langle \left(2-\cos\varphi_{1}\right)\left(1-\cos\varphi_{2}\right)\right\rangle \\
 &  & +_{*}\includegraphics[width=0.5cm]{graph-b4-sgm3aI.pdf}_{*}\,\left(1+2\left\langle 1-\cos\varphi_{1}\right\rangle +\left\langle 1-\cos\varphi_{1}-\cos\varphi_{2}\right\rangle \right)\Bigr]
\end{eqnarray*}
Now, we replace this result in Eq. \eqref{eq:B4-3a} to obtain
\begin{equation}
\includegraphics[width=0.5cm]{graph-b4-fs.pdf}=-\frac{\sigma^{3d}}{d^{3}}\Bigl[\,\includegraphics[width=0.5cm]{graph-b4-sgm3I.pdf}\left\langle 4-2\cos\varphi\right\rangle +\includegraphics[width=0.5cm]{graph-b4-sgm3aI.pdf}4\left\langle \left(2-\cos\varphi_{1}\right)\left(2-\cos\varphi_{2}\right)-1\right\rangle \Bigr]\:.\label{eq:B4s-resD3}
\end{equation}
Here, the stars are unnecessary and thus they were not written. This
expression is one of the results of the present work, once the angular
variables related with rigid rotations are integrated on the remaining
are two three-folded angular integrals.

\section{Integration of the complete star\label{sec:The-three-foldedIntegral}}

The star integral is thus
\begin{equation}
\includegraphics[width=0.5cm]{graph-b4-fs.pdf}=A\left(Y+U\right)\:,\label{eq:AYU}
\end{equation}
where 
\begin{equation}
A=\frac{\sigma^{3d}}{d^{3}}\Omega_{d}\Omega_{d-1}\Omega_{d-2}\:\textrm{ with }\:\Omega_{d}=\frac{d\pi^{d/2}}{\Gamma(d/2+1)}=\frac{2dB_{2}}{\sigma^{d}}\:,\label{eq:A}
\end{equation}
 being $\Omega_{d}$ the surface area of the unit sphere in the $d$
dimensional space (with $\Omega_{k}=1$ if $k<1$). The term related
with \includegraphics[clip,width=0.5cm,height=0.5cm]{graph-b4-sgm3I}
is
\begin{equation}
Y=\int_{0}^{\frac{\pi}{3}}\int_{0}^{\frac{\pi}{3}}\int_{0}^{\theta_{max,\textrm{I}}}\!\!h_{\textrm{I}}\sin\left(\varphi_{1}\right)^{d-2}\sin\left(\varphi_{2}\right)^{d-2}\sin\left(\theta\right)^{d-3}d\theta d\varphi_{1}d\varphi_{2}\:,\label{eq:intY00}
\end{equation}
where $h_{\textrm{I}}=4-2\cos\varphi_{1}$ and $\theta_{max,\textrm{I}}=\min\left\{ \arccos\left(\frac{\frac{1}{2}-\cos\varphi_{1}\cos\varphi_{2}}{\sin\varphi_{1}\sin\varphi_{2}}\right),\pi\right\} $.
A second form for $Y$ is obtained through a change of variable. Let
be $\tan\psi=\tan\varphi_{2}\cos\theta$, $\cos\vartheta=\sin\varphi_{2}\sin\theta$
(and let rename $\varphi=\varphi_{1}$), the Eq. \eqref{eq:intY00}
transforms into the simpler expression
\begin{equation}
Y=2\int_{0}^{\frac{\pi}{3}}\int_{\frac{\varphi}{2}}^{\frac{\pi}{3}}\int_{\vartheta_{min}}^{\frac{\pi}{2}}\!h_{\textrm{I}}\sin\left(\varphi\right)^{d-2}\sin\vartheta\cos\left(\vartheta\right)^{d-3}d\vartheta\,d\psi\,d\varphi\:,\label{eq:intY01}
\end{equation}
with $\vartheta_{min}=\arccos\left(\sqrt{1-\left(\frac{1}{2}\sec\psi\right)^{2}}\right)$,
here the integral in $\vartheta$ is trivial. The term related with
\includegraphics[clip,width=0.4cm,height=0.4cm]{graph-b4-sgm3aI}
is
\begin{equation}
U=\int_{0}^{\frac{\pi}{3}}\int_{0}^{\frac{\pi}{3}}\int_{0}^{\theta_{max,\textrm{II}}}\!\!h_{\textrm{II}}\sin\left(\varphi_{1}\right)^{d-2}\sin\left(\varphi_{2}\right)^{d-2}\sin\left(\theta\right)^{d-3}\,d\theta\,d\varphi_{1}d\varphi_{2}\:,\label{eq:intU00}
\end{equation}
with $h_{\textrm{II}}=4\left[\left(2-\cos\varphi_{1}\right)\left(2-\cos\varphi_{2}\right)-1\right]$
and $\theta_{max,\textrm{II}}=\arccos\left[\tan\frac{\varphi_{1}}{2}\tan\frac{\varphi_{2}}{2}\right]$.
Here the integral in $\theta$ gives $B_{1-\left(\tan\frac{\varphi_{1}}{2}\tan\frac{\varphi_{2}}{2}\right)^{2}}\left(\frac{d}{2}-1,\frac{1}{2}\right)$.
Thus, for arbitrary $d\geq3$ the Eq. \eqref{eq:AYU} transforms to
\begin{eqnarray}
\includegraphics[width=0.5cm]{graph-b4-fs.pdf} & = & \frac{2A}{d-2}\int_{0}^{\frac{\pi}{3}}\int_{\frac{\varphi}{2}}^{\frac{\pi}{3}}h_{\textrm{I}}\sin\left(\varphi\right)^{d-2}\left[1-\left(\frac{\sec\psi}{2}\right)^{2}\right]^{\frac{d}{2}-1}d\psi d\varphi+\nonumber \\
 &  & \frac{A}{2}\int_{0}^{\frac{\pi}{3}}\int_{0}^{\frac{\pi}{3}}h_{\textrm{II}}\sin\left(\varphi_{1}\right)^{d-2}\sin\left(\varphi_{2}\right)^{d-2}B_{1-\left(\tan\frac{\varphi_{1}}{2}\tan\frac{\varphi_{2}}{2}\right)^{2}}\left(\frac{d}{2}-1,\frac{1}{2}\right)\,d\varphi_{1}d\varphi_{2}\:,\label{eq:B4d_fnal}
\end{eqnarray}
Last equation and Eqs. (\ref{eq:AYU}, \ref{eq:intY01}, \ref{eq:intU00})
are convenient for both numerical and analytical integration, even
for non-integer $d$. In the following we focus on even dimension
$d=2m+4$ with $m$ any non-negative integer, i.e, even $d\geq4$.
The rest of this Section is devoted to evaluate $Y$ and $U$.

\subsection*{Integration of $Y$ for even dimensions}

We expand the binomial $\left[1-\left(\frac{\sec\psi}{2}\right)^{2}\right]^{m+1}$
to obtain
\begin{equation}
Y=\frac{1}{m+1}\int_{0}^{\frac{\pi}{3}}\int_{\frac{\varphi}{2}}^{\frac{\pi}{3}}\!h_{I}\sin\left(\varphi\right)^{2m+2}\left[1+\sum_{n=1}^{m+1}\binom{m+1}{n}\left(-1\right)^{n}4^{-n}\cos\left(\psi\right)^{-2n}\right]d\psi d\varphi\:.\label{eq:intY}
\end{equation}
Next, we solve the integral in $\psi$ variable. Let $n$ be any positive
integer, according to\footnote{Ref. \citep{GradshteynRyzhik2007}, Sec. 2.519 Eq.(1), at p.156.}
\begin{equation}
\int\cos\left(x\right)^{-2n}\,dx=\frac{\sin x}{2n-1}\Biggl[\cos\left(x\right)^{-2n+1}+\sum_{k=1}^{n-1}\frac{2^{k}\prod_{i=1}^{k}(n-i)}{\prod_{i=1}^{k}(2n-2i-1)}\cos\left(x\right)^{-2n+2k+1}\Biggr]\:.\label{eq:Cxm2n}
\end{equation}
One finds
\begin{eqnarray}
\int_{\frac{\varphi}{2}}^{\frac{\pi}{3}}\cos\left(\psi\right)^{-2n}d\psi & = & \frac{1}{2}B_{\cos\left(\frac{\varphi}{2}\right)^{2}}\left(\frac{1}{2}-n,\frac{1}{2}\right)-\frac{1}{2}B_{\frac{1}{4}}\left(\frac{1}{2}-n,\frac{1}{2}\right)\label{eq:Cxm2n-1}\\
 & = & \sqrt{3}C-\frac{\sin\frac{\varphi}{2}}{2n-1}\Biggl[\cos\left(\frac{\varphi}{2}\right)^{-2n+1}+\sum_{k=1}^{n-1}\frac{2^{k}\prod_{i=1}^{k}(n-i)}{\prod_{i=1}^{k}(2n-2i-1)}\cos\left(\frac{\varphi}{2}\right)^{-2n+2k+1}\Biggr]\:,
\end{eqnarray}
here $C=\frac{4^{n-1}}{2n-1}\left(1+\sum_{k=1}^{n-1}\frac{2^{k}\prod_{i=1}^{k}(n-i)}{4^{k}\prod_{i=1}^{k}(2n-2i-1)}\right)$.
We introduce the shortcut $K=4\sum_{n=1}^{m+1}\binom{m+1}{n}(-4)^{-n}C$,
i.e. $K=\sum_{n=1}^{m+1}\binom{m+1}{n}\frac{(-1)^{n}}{2n-1}\left[1+\sum_{k=1}^{n-1}\frac{\prod_{i=1}^{k}(n-i)}{2^{k}\prod_{i=1}^{k}(2n-2i-1)}\right]$,
to obtain,
\begin{eqnarray}
\int_{\frac{\varphi}{2}}^{\frac{\pi}{3}}\int_{\theta_{min}}^{\frac{\pi}{2}}\!\sin\vartheta\cos\left(\vartheta\right)^{2m+1}d\psi d\vartheta & = & \frac{2^{-1}}{m+1}\left(\frac{\sqrt{3}}{4}K+\frac{\pi}{3}-\frac{\varphi}{2}\right)-\frac{2^{-1}}{m+1}\sum_{n=1}^{m+1}\binom{m+1}{n}\frac{\left(-1\right)^{n}4^{-n}}{2n-1}\times\nonumber \\
 &  & \sin\frac{\varphi}{2}\Biggl[\cos\left(\frac{\varphi}{2}\right)^{-2n+1}+\sum_{k=1}^{n-1}\frac{2^{k}\prod_{i=1}^{k}(n-i)}{\prod_{i=1}^{k}(2n-1-2i)}\cos\left(\frac{\varphi}{2}\right)^{-2n+2k+1}\Biggr]\:.\label{eq:intY02}
\end{eqnarray}
To solve the integral in $\varphi$ we separate the linear term $\frac{\sqrt{3}}{4}K+\frac{\pi}{3}-\frac{\varphi}{2}$
from that involving $\sin\frac{\varphi}{2}$ and $\cos\frac{\varphi}{2}$.
$L\equiv\int_{0}^{\frac{\pi}{3}}\left(\frac{\sqrt{3}}{4}K+\frac{\pi}{3}-\frac{\varphi}{2}\right)\left(4-2\cos\varphi\right)\sin\left(\varphi\right)^{2m+2}d\varphi$
splits in four integrals: $\left(\sqrt{3}K+\frac{4\pi}{3}\right)\int_{0}^{\frac{\pi}{3}}\sin\left(\varphi\right)^{2m+2}d\varphi$,
$-\left(\frac{\sqrt{3}}{2}K+\frac{2\pi}{3}\right)\times$ $\int_{0}^{\frac{\pi}{3}}\cos\varphi\sin\left(\varphi\right)^{2m+2}d\varphi$,
$-2\int_{0}^{\frac{\pi}{3}}\varphi\sin\left(\varphi\right)^{2m+2}d\varphi$,
and $\int_{0}^{\frac{\pi}{3}}\varphi\cos\varphi\sin\left(\varphi\right)^{2m+2}d\varphi$.
Each of them is solved separately in Appendix \ref{Appsec:IntegL}.
The result is
\begin{eqnarray}
L & = & \frac{\pi^{2}(2m+1)!!}{3\,2^{m+1}(m+1)!}+\frac{\pi\sqrt{3}}{6}\left[\frac{(2m+1)!!K}{2^{m}(m+1)!}-Q_{0}-\frac{\left(\frac{3}{4}\right)^{m+1}}{2m+3}\right]\nonumber \\
 &  & -\frac{2^{m+1}(m+1)!}{(2m+3)^{2}(2m+1)!!}-\frac{3}{4}K\left[Q_{0}+\frac{\left(\frac{3}{4}\right)^{m+1}}{2m+3}\right]-Q_{1}+\frac{Q_{2}}{(2m+3)^{2}}\:,\label{eq:Lres}
\end{eqnarray}
with 
\begin{eqnarray}
K & = & \sum_{n=1}^{m+1}(-1)^{n}\binom{m+1}{n}\sum_{k=0}^{n-1}\frac{\prod_{i=1}^{k}(n-i)}{2^{k}\prod_{i=0}^{k}(2n-2i-1)}\:,\nonumber \\
Q_{0} & = & \sum_{k=0}^{m}\frac{\left(\frac{3}{4}\right)^{m-k}\prod_{i=0}^{k-1}(2m-2i+1)}{2^{k+1}\prod_{i=0}^{k}(m-i+1)}\:,\nonumber \\
Q_{1} & = & \sum_{k=0}^{m}\frac{\left(\frac{3}{4}\right)^{m-k+1}\prod_{i=0}^{k-1}(2m-2i+1)}{2^{k+1}(m-k+1)\prod_{i=0}^{k}(m-i+1)}\:,\label{eq:KQ}\\
Q_{2} & = & \sum_{k=0}^{m+1}\frac{\left(\frac{3}{4}\right)^{m-k+1}2^{k-1}\prod_{i=0}^{k-1}(m-i+1)}{\prod_{i=0}^{k-1}(2m-2i+1)}\:.\nonumber 
\end{eqnarray}
Here $K$, $Q_{0}$, $Q_{1}$ and $Q_{2}$ are rational numbers.
Besides, to evaluate $L$ for a given even dimension one has to add
$\sim2d$ terms.

Now, we focus on the integration of the term including trigonometric
functions of $\frac{\varphi}{2}$. Using the shortcut $\phi=\frac{\varphi}{2}$,
the terms are of the type $\int_{0}^{\frac{\pi}{6}}2h(2\sin\phi\cos\phi)^{2m+2}\sin\phi\cos^{-2n+1}\phi\,d\phi$
with $h=2+4\sin^{2}\phi$. Thus, we need to solve integrals like $\int_{0}^{\frac{\pi}{6}}\cos^{2l+1}\phi\sin^{2n+1}\phi\,d\phi$.

Let us assume $p$ and $q$ be non-negative integers to introduce
$D\left(p,q\right)\equiv\int_{0}^{\frac{\pi}{6}}\sin\left(\phi\right)^{p}\cos\left(\phi\right)^{q}d\phi=\frac{1}{2}B_{\frac{1}{4}}\left(\frac{p+1}{2},\frac{q+1}{2}\right)$
$=\frac{1}{2}B\left(\frac{p+1}{2},\frac{q+1}{2}\right)-\frac{1}{2}B_{\frac{3}{4}}\left(\frac{q+1}{2},\frac{p+1}{2}\right)$.
Let $n$ be any positive integer and $q\in\mathbb{R}_{-\left\{ -1,-3,\ldots,-(2n-1)\right\} }$.
According to \footnote{Ref. \citep{GradshteynRyzhik2007}, Sec. 2.512 Eq.(4), at p.152.}
we have the identity
\begin{equation}
\int_{0}^{u}\sin\left(x\right)^{2n+1}\cos\left(x\right)^{q}dx=\frac{2^{n}n!\left(q-1\right)!!}{\left(2n+q+1\right)!!}-\frac{\cos\left(u\right)^{q+1}}{2n+q+1}\left[\sin\left(u\right)^{2n}+\sum_{k=1}^{n}\frac{2^{k}\prod_{i=0}^{k-1}(n-i)}{\prod_{i=0}^{k-1}(2n+q-2i-1)}\sin\left(u\right)^{2n-2k}\right]\,.\label{eq:intSnxiCsxq}
\end{equation}
For $q=2l+1$, with $n$ and $l$ positive integers, we obtain
\begin{equation}
D\left(2n+1,2l+1\right)=\frac{n!\,l!}{2(n+l+1)!}-\frac{\left(\frac{3}{4}\right)^{l+1}}{4^{n}2(n+l+1)}\left[1+\sum_{k=1}^{n}\frac{4^{k}\prod_{i=0}^{k-1}(n-i)}{\prod_{i=0}^{k-1}(n+l-i)}\right]\,.\label{eq:DimparX2}
\end{equation}
Let $n$ be any positive integer and $q\in\mathbb{R}_{-\left\{ -2,-4,\ldots,-2n\right\} }$.
According to \footnote{Ref. \citep{GradshteynRyzhik2007}, Sec. 2.512 Eqs.(1,2), at p.152.}
we have
\begin{eqnarray}
\int_{0}^{u}\sin\left(x\right)^{2n}\cos\left(x\right)^{q}dx & = & -\frac{\cos^{q+1}u}{2n+q}\left[\sin\left(u\right)^{2n-1}+\sum_{k=1}^{n-1}\frac{\prod_{i=1}^{k}\left(2n-2i+1\right)}{\prod_{i=1}^{k}\left(2n+2l-2i\right)}\sin\left(u\right)^{2n-2k-1}\right]\nonumber \\
 &  & +\frac{\left(2n-1\right)!!q!!}{\left(2n+q\right)!!}\int_{0}^{u}\cos\left(x\right)^{q}dx\,,\label{eq:Snx2nCsxp}\\
\int_{0}^{u}\cos\left(x\right)^{2l}dx & = & u\frac{\left(2l-1\right)!!}{2^{l}l!}+\frac{\sin u}{2l}\left[\cos\left(u\right)^{2l-1}+\sum_{k=1}^{l-1}\frac{\prod_{i=1}^{k}\left(2l-2i+1\right)}{2^{k}\prod_{i=1}^{k}\left(l-i\right)}\cos\left(u\right)^{2l-2k-1}\right]\,.\label{eq:Csx2n}
\end{eqnarray}
Thus, for $q=2l$, with $n$ and $l$ positive integers we obtain
\begin{eqnarray}
D\left(2n,2l\right) & = & \frac{\sqrt{3}\,3^{l-1}(l-1)!(2n-1)\text{!!}}{2^{2l+n+1}(l+n)!}\left[1+\sum_{k=1}^{l-1}\frac{\left(\frac{2}{3}\right)^{k}\prod_{i=1}^{k}(2l-2i+1)}{\prod_{i=1}^{k}(l-i)}\right]\nonumber \\
 &  & +\frac{\pi(2l-1)\text{!!}(2n-1)\text{!!}}{2^{l+n}6\,(l+n)!}-\frac{\sqrt{3}\,3^{l}}{2^{2l+2n+1}(l+n)}\left[1+\sum_{k=1}^{n-1}\frac{2^{k}\prod_{i=1}^{k}(2n-2i+1)}{\prod_{i=1}^{k}(l+n-i)}\right]\,.\label{eq:DparX2}
\end{eqnarray}
To evaluate terms like $\int_{0}^{\frac{\pi}{6}}2h\sin\left(2\phi\right)^{2m+2}\sin\left(\phi\right)\cos\left(\phi\right)^{-2l+1}d\phi$
we introduce
\begin{eqnarray}
g_{\mathrm{1}}\left(m,l\right) & \equiv & 4^{-m-2}\int_{0}^{\frac{\pi}{6}}2h\sin\left(2\phi\right)^{2m+2}\sin\left(\phi\right)\cos\left(\phi\right)^{-2l+1}d\phi\nonumber \\
 & = & \int_{0}^{\frac{\pi}{6}}\left(1+2\sin\left(\phi\right)^{2}\right)\sin\left(\phi\right)^{2m+3}\cos\left(\phi\right)^{2m-2l+3}d\phi\nonumber \\
 & = & D\left(2m+3,2m-2l+3\right)+2D\left(2m+5,2m-2l+3\right)\:.\label{eq:g1}
\end{eqnarray}
Collecting the partial results, we find
\begin{equation}
Y=\frac{L}{m+1}-\frac{M}{m+1}\:,\label{eq:Yres}
\end{equation}
with $L$ given in Eqs. (\ref{eq:Lres}, \ref{eq:KQ}) and
\begin{equation}
M=4^{2}\sum_{n=1}^{m+1}\binom{m+1}{n}\left(-1\right)^{n}4^{m-n}{\displaystyle \sum_{k=0}^{n-1}}\frac{2^{k}\prod_{i=1}^{k}(n-i)}{\prod_{i=0}^{k}(2n-2i-1)}g_{\mathrm{1}}(m,n-k)\;.\label{eq:M}
\end{equation}
To evaluate $Y$ for a given large even dimension one has to add
order $\sim\frac{d^{3}}{8}$ different terms.

\subsection*{Integration of $U$ for even dimensions}

To solve the first integral in
\begin{equation}
U\equiv\int_{0}^{\frac{\pi}{3}}\int_{0}^{\frac{\pi}{3}}\int_{0}^{\theta_{max}}h\sin\left(\varphi_{1}\right)^{2m+2}\sin\left(\varphi_{2}\right)^{2m+2}\sin\left(\theta\right)^{2m+1}d\theta\,d\varphi_{1}d\varphi_{2}\;,\label{eq:intU01}
\end{equation}
with $h=4\left[\left(2-\cos\varphi_{1}\right)\left(2-\cos\varphi_{2}\right)-1\right]$
and $\theta_{max}=\arccos\left[\tan\frac{\varphi_{1}}{2}\tan\frac{\varphi_{2}}{2}\right]$,
we use the identity \footnote{Ref. \citep{GradshteynRyzhik2007}, Sec. 2.511 Eq.(3), at p.152.}
\begin{equation}
\int_{0}^{u}\sin\left(x\right)^{2n+1}dx=\frac{2^{n}n!}{\left(2n+1\right)!!}-\frac{\cos u}{2n+1}\left[\sin\left(u\right)^{2n}+\sum_{k=0}^{n-1}\frac{2^{k+1}n!}{(n-k-1)!}\frac{\left(2n-2k-3\right)!!}{\left(2n-1\right)!!}\sin\left(u\right)^{2n-2k-2}\right]\,.\label{eq:Snx2np1}
\end{equation}
The case $m=0$ can be solved without using Eq. \eqref{eq:Snx2np1}.
We introduce the shortcut $\phi_{1}=\varphi_{1}/2$ and $\phi_{2}=\varphi_{2}/2$
to express $\cos\theta_{max}\left(\sin\theta_{max}\right)^{2n}=\tan\phi_{1}\tan\phi_{2}\left(1-\tan\phi_{1}^{2}\tan\phi_{2}^{2}\right)^{n}=\sum_{i=0}^{n}\binom{n}{i}\left(-1\right)^{i}\left(\tan\phi_{1}\tan\phi_{2}\right)^{2i+1}$.
Thus, for $m$ any positive integer we obtain,
\begin{eqnarray}
 &  & \int_{0}^{\theta_{max}}\sin\left(\theta\right)^{2m+1}d\theta=C-\frac{1}{2m+1}\sum_{n=0}^{m}\binom{m}{n}\left(-1\right)^{n}\left(\tan\phi_{1}\tan\phi_{2}\right)^{2n+1}\nonumber \\
 &  & -\frac{1}{2m+1}\left[\sum_{k=0}^{m-1}\frac{2^{k+1}\prod_{i=0}^{k}(m-i)}{\prod_{i=0}^{k}(2m-2i-1)}\sum_{n=0}^{m-k-1}\binom{m-k-1}{n}\left(-1\right)^{n}\left(\tan\phi_{1}\tan\phi_{2}\right)^{2n+1}\right]\:,\label{eq:intU02}
\end{eqnarray}
here $C=\frac{2^{m}m!}{\left(2m+1\right)!!}$. This expression must
be replaced in Eq. \eqref{eq:intU01}. To complete the integration
in $\varphi_{1}$ and $\varphi_{2}$ for each term at the right of
Eq. \eqref{eq:intU02} we transform $h$ to $h=16\sin^{2}\phi_{1}+16\sin^{2}\phi_{1}\sin^{2}\phi_{2}$.
The complete integral of the constant term $C$ is
\begin{eqnarray}
C\int_{0}^{\frac{\pi}{3}}\int_{0}^{\frac{\pi}{3}}h\sin\left(\varphi_{1}\right)^{2m+2}\sin\left(\varphi_{2}\right)^{2m+2}d\varphi_{1}d\varphi_{2} & = & \frac{2^{5m+10}m!}{\left(2m+1\right)!!}\int_{0}^{\frac{\pi}{6}}\int_{0}^{\frac{\pi}{6}}\left[\left(\sin\phi_{1}\right)^{2}+\left(\sin\phi_{1}\sin\phi_{2}\right)^{2}\right]\times\nonumber \\
 &  & \left(\sin\phi_{1}\cos\phi_{1}\sin\phi_{2}\cos\phi_{2}\right)^{2m+2}\,d\phi_{1}d\phi_{2}\:.\label{eq:intU02a}
\end{eqnarray}
It includes terms of the form $D\left(2n,2l\right)$ given in Eq.
\eqref{eq:DparX2}. We introduce the shortcut
\begin{equation}
g_{\mathrm{2}}\left(m\right)=D\left(2m+4,2m+2\right)\left[D\left(2m+4,2m+2\right)+D\left(2m+2,2m+2\right)\right]\,,\label{eq:Gc}
\end{equation}
and thus, the r.h.s of Eq. \eqref{eq:intU02a} reduces to
\begin{equation}
\frac{2^{5m+10}m!}{\left(2m+1\right)!!}g_{\mathrm{2}}\left(m\right)\,.\label{eq:intU02c}
\end{equation}

The remaining terms in Eqs. (\ref{eq:intU01}, \ref{eq:intU02}) contain
\begin{eqnarray}
 &  & \int_{0}^{\frac{\pi}{3}}\!\int_{0}^{\frac{\pi}{3}}h\left(\tan\phi_{1}\tan\phi_{2}\right)^{2i+1}\sin\left(\varphi_{1}\right)^{2m+2}\sin\left(\varphi_{2}\right)^{2m+2}d\varphi_{1}d\varphi_{2}=\nonumber \\
 &  & =2^{4m+10}\int_{0}^{\frac{\pi}{6}}\!\int_{0}^{\frac{\pi}{6}}\left(\sin\phi_{1}^{2}+\sin\phi_{1}^{2}\sin\phi_{2}^{2}\right)\left(\tan\phi_{1}\tan\phi_{2}\right)^{2i+1}\times\nonumber \\
 &  & \qquad\left(\sin\phi_{1}\cos\phi_{1}\sin\phi_{2}\cos\phi_{2}\right)^{2m+2}\,d\phi_{1}d\phi_{2}\:,\label{eq:intU03}
\end{eqnarray}
which involves terms of type $D\!\left(2n+1,2l+1\right)$ analyzed
in Eq. \eqref{eq:DimparX2}. Therefore Eq. \eqref{eq:intU03} gives
$2^{2m+10}g_{\mathrm{3}}\left(m,i\right)$ with
\begin{eqnarray}
 &  & g_{\mathrm{3}}\left(m,i\right)=D\!\bigl[2\left(m+i+2\right)+1,2m-2i+1\bigr]\times\nonumber \\
 &  & \biggl[D\!\bigl[2\left(m+i+2\right)+1,2m-2i+1\bigr]+D\!\bigl[2\left(m+i+1\right)+1,2m-2i+1\bigr]\biggr]\:,\label{eq:g3}
\end{eqnarray}
which is also a rational number.

Collecting the partial results, we obtain
\begin{eqnarray}
U & = & \frac{2^{5(m+2)}m!}{(2m+1)\text{!!}}g_{\mathrm{2}}\left(m\right)-\frac{4^{2m+5}}{2m+1}\sum_{k=0}^{m}\left[\frac{2^{k}\prod_{j=0}^{k-1}(m-j)}{\prod_{j=0}^{k-1}(2m-2j-1)}\sum_{i=0}^{m-k}\binom{m-k}{i}\left(-1\right)^{i}g_{\mathrm{3}}\left(m,i\right)\right]\:.\label{eq:intUresult}
\end{eqnarray}
It is interesting to note that the factors $\sqrt{3}$ and $\pi$
originated in this expression come from $g_{\mathrm{2}}\left(m\right)$,
while the second term on the right of Eq. \eqref{eq:intUresult} is
a rational number. Moreover, all the factors $\sqrt{3}$ and $\pi$
appearing in $Y$ are those explicitly written in Eq. \eqref{eq:Lres}.
Naturally, $A$ in Eq. \eqref{eq:AYU} includes powers of $\pi$ originated
in the solid angles, in fact $A\sim\pi^{-2}$.

\section{Results\label{sec:Results}}

The main result we obtained, is the explicit expression of \includegraphics[width=0.4cm]{graph-b4-fs}
valid for even dimensions $d>2$, given in Eqs. (\ref{eq:AYU}, \ref{eq:Lres},
\ref{eq:Yres}) and \eqref{eq:intUresult}. Once it is complemented
with Eqs. (\ref{eq:B4MayerDiag}, \ref{eq:B4SqrE}) and \eqref{eq:B4SqrdE}
one finds $B_{4}(d)$. For even $d=2\nu$ (positive integer $\nu\geq2$),
we derive the following form for $B_{4}:$
\begin{equation}
\frac{B_{4}}{B_{2}^{3}}=2-\frac{\sqrt{3}}{\pi}a_{4,1}+\frac{1}{\pi^{2}}a_{4,2}\:.\label{eq:FrB4}
\end{equation}
Here, $a_{4,1}$ and $a_{4,2}$ are positive rational numbers (we
note that $\underset{\nu\rightarrow\infty}{\lim}\frac{\sqrt{3}}{\pi}a_{4,1}(\nu)=6$
and $\underset{\nu\rightarrow\infty}{\lim}\pi^{-2}a_{4,2}(\nu)=4$).
In fact, we verify that
\begin{equation}
a_{4,1}=3\sum_{i=1}^{\nu}\frac{\left(\frac{3}{2}\right)^{i}(i-1)!}{(2i-1)\text{!!}}\:.\label{eq:a41}
\end{equation}
Unfortunately, we do not find a simple expression for $a_{4,2}$.
Instead, we obtain

\begin{eqnarray}
a_{4,2} & = & \left(\frac{(d-2)!!}{(d-1)!!}\right)^{2}+\frac{9}{2}R_{0}^{2}-3R_{1}+3R_{3}+\frac{(d-2)!!}{(d-3)!!}\Biggl[-\frac{24\,d}{d-1}R_{2}\nonumber \\
 &  & +K\left(\frac{2^{-d}3^{d/2}}{d-1}+\frac{3}{4}Q_{0}\right)+Q_{1}-\frac{Q_{2}}{(d-1)^{2}}+M-\left(\frac{d}{2}-1\right)(U_{1}+U_{2})\Biggr]\:,\label{eq:a42}
\end{eqnarray}
with $R_{0}$, $R_{1}$ and $R_{2}$ given in Eq. \eqref{eq:R123},
\begin{equation}
R_{3}=\sum_{k=1}^{\nu}\frac{(k-1)!2^{k-1}}{k(2k-1)!!}\biggl[4+\frac{\nu\,(2\nu)!!\,4^{k}k!\,(\nu+k-1)!}{(2\nu-1)!!\,(\nu+2k)!}\biggr]\:,\label{eq:R3}
\end{equation}
$K$, $Q_{0}$, $Q_{1}$ and $Q_{2}$ given in Eq. \eqref{eq:KQ},
and 
\begin{eqnarray}
M & = & 4^{2}\sum_{n=1}^{m+1}\binom{m+1}{n}\left(-1\right)^{n}4^{m-n}{\displaystyle \sum_{k=0}^{n-1}}\Biggl\{\frac{2^{k}\prod_{i=1}^{k}(n-i)}{\prod_{i=0}^{k}(2n-2i-1)}\times\nonumber \\
 &  & \Bigl[D[2m+3,2m-2(n-k)+3]+2D[2m+5,2m-2(n-k)+3]\Bigr]\Biggr\}\:,\nonumber \\
U_{1} & = & \frac{3^{2m+2}m!}{2^{3m+2}(m+1)^{2}(2m+1)!!}\times\nonumber \\
 &  & \left[\frac{3m^{2}+7m+3}{2m+3}-\sum_{k=1}^{m}\frac{\left(\frac{2}{3}\right)^{k}\prod_{i=1}^{k}(2m-2i+3)}{\prod_{i=1}^{k}(m-i+1)}+\frac{3}{4}\sum_{k=1}^{m}\frac{(2m-k+2)2^{k}\prod_{i=1}^{k}(2m-2i+3)}{(2m-2k+3)\prod_{i=1}^{k}(2m-i+2)}\right]\times\nonumber \\
 &  & \left[\frac{(m+1)(m+2)}{2m+3}-\sum_{k=1}^{m}\frac{\left(\frac{2}{3}\right)^{k}\prod_{i=1}^{k}(2m-2i+3)}{\prod_{i=1}^{k}(m-i+1)}+\sum_{k=1}^{m}\frac{\left(1+\frac{2m-k+2}{4(2m-2k+3)}\right)2^{k}\prod_{i=1}^{k}(-2i+2m+3)}{\prod_{i=1}^{k}(2m-i+2)}\right]\:,\label{eq:MU1U2}\\
U_{2} & = & -\frac{4^{2m+5}}{2m+1}\sum_{k=0}^{m}\Biggl\{\frac{2^{k}\prod_{j=0}^{k-1}(m-j)}{\prod_{j=0}^{k-1}(2m-2j-1)}\sum_{i=0}^{m-k}\binom{m-k}{i}\left(-1\right)^{i}\times\nonumber \\
 &  & \Bigl[D[2(m+i+2)+1,2m-2i+1]+D[2(m+i+1)+1,2m-2i+1]\Bigr]\Biggr\}\:,\nonumber 
\end{eqnarray}
where the function $D(a,b)$ for $a$ and $b$ odd numbers is defined
in Eq. \eqref{eq:DimparX2}.

In Table \ref{tab:B4} the exact values of $B_{4}$ for $d\leq50$
are shown, there the new results correspond to even dimensions between
$d=14$ and $d=50$. As it is usual, results are presented as the
ratio $B_{4}/B_{2}^{3}$. The known values for even and odd dimensions
$d\leq12$ are included. Additionally, we have evaluated $B_{4}$
for each even dimension up to $d=200$ and also for several higher
dimensions up to $d=1000$. Because their length, which increase with
dimension, the obtained expressions are not given here. In the Supplementary
Material, Sec. \ref{sec:SupplMat}, we give the exact value of $B_{4}/B_{2}^{3}$
for $d=100$, $200$, $300$, $500$ and $1000$. Also, a computer
algebra program that allows to evaluate $B_{4}$ is included.
\begin{table}[H]
\begin{centering}
\begin{tabular}{|c|c|cc|}
\hline 
$d$ & $B_{4}/B_{2}^{3}$ & decimal expansion & \tabularnewline
\hline 
1 & $1$ & $1$ & \tabularnewline
2 & $2-\frac{9\sqrt{3}}{2\pi}+\frac{10}{\pi^{2}}$ & $0.53223180702...$ & \citep{Rowlinson_1964_b,Hemmer_1965}\tabularnewline
3 & $\frac{2707}{4480}+\frac{219}{2240}\frac{\sqrt{2}}{\pi}-\frac{4131}{4480}\frac{\arccos(1/3)}{\pi}$ & $0.28694950598...$ & \citep{vanLaar_1899_b}\tabularnewline
4 & $2-\frac{27\sqrt{3}}{4\pi}+\frac{832}{45\pi^{2}}$ & $0.15184606235...$ & \citep{Clisby_2004}\tabularnewline
5 & $\frac{25315393}{32800768}+\frac{3888425}{16400384}\frac{\sqrt{2}}{\pi}-\frac{67183425}{32800768}\frac{\arccos(1/3)}{\pi}$ & $0.07597248028...$ & \citep{Lyberg_2005}\tabularnewline
6 & $2-\frac{81\sqrt{3}}{10\pi}+\frac{38848}{1575\pi^{2}}$ & $0.033363148440...$ & \citep{Clisby_2004}\tabularnewline
7 & $\frac{299189248759}{290596061184}+\frac{159966456685}{435894091776}\frac{\sqrt{2}}{\pi}-\frac{292926667005}{96865353728}\frac{\arccos(1/3)}{\pi}$ & $0.00986494662...$ & \citep{Lyberg_2005}\tabularnewline
8 & $2-\frac{2511\sqrt{3}}{280\pi}+\frac{17605024}{606375\pi^{2}}$ & $-0.0025576848788...$ & \citep{Clisby_2004}\tabularnewline
9 & $\frac{2886207717678787}{2281372811001856}+\frac{2698457589952103}{5703432027504640}\frac{\sqrt{2}}{\pi}-\frac{8656066770083523}{2281372811001856}\frac{\arccos(1/3)}{\pi}$ & $-0.00858079817...$ & \citep{Lyberg_2005}\tabularnewline
10 & $2-\frac{2673\sqrt{3}}{280\pi}+\frac{49048616}{1528065\pi^{2}}$ & $-0.010962474302...$ & \citep{Clisby_2004}\tabularnewline
11 & $\frac{17357449486516274011}{11932824186709344256}+\frac{16554115383300832799}{29832060466773360640}\frac{\sqrt{2}}{\pi}-\frac{52251492946866520923}{11932824186709344256}\frac{\arccos(1/3)}{\pi}$ & $-0.0113371986...$ & \citep{Lyberg_2005}\tabularnewline
12 & $2-\frac{2187\sqrt{3}}{220\pi}+\frac{11565604768}{337702365\pi^{2}}$ & $-0.010670280553...$ & \citep{Clisby_2004}\tabularnewline
\hline 
14 & $2-\frac{37179\sqrt{3}}{3640\pi}+\frac{4687062562736}{131076760815\pi^{2}}$ & $-8.2212455146...\!\times\!10^{-3}$ & \tabularnewline
16 & $2-\frac{833247\sqrt{3}}{80080\pi}+\frac{18316976338530304}{497436307292925\pi^{2}}$ & $-5.7523349690...\!\times\!10^{-3}$ & \tabularnewline
18 & $2-\frac{14348907\sqrt{3}}{1361360\pi}+\frac{34423898313743697664}{916111865931136875\pi^{2}}$ & $-3.8229418415...\!\times\!10^{-3}$ & \tabularnewline
\hline 
20 & $2-\frac{275109291\sqrt{3}}{25865840\pi}+\frac{3847541019771123277312}{100955527625611283625\pi^{2}}$ & $-2.4621254986...\!\times\!10^{-3}$ & \tabularnewline
22 & $2-\frac{276880761\sqrt{3}}{25865840\pi}+\frac{10951650004732725201152}{284511032399449981125\pi^{2}}$ & $-1.5530223910...\!\times\!10^{-3}$ & \tabularnewline
24 & $2-\frac{246057183\sqrt{3}}{22881320\pi}+\frac{1040943676896678306727936}{26851955023354986149625\pi^{2}}$ & $-9.6537457745...\!\times\!10^{-4}$ & \tabularnewline
26 & $2-\frac{8023164777\sqrt{3}}{743642900\pi}+\frac{8203500529370190905316494848}{210550310408130039474001875\pi^{2}}$ & $-5.9370020421...\!\times\!10^{-4}$ & \tabularnewline
28 & $2-\frac{32168655171\sqrt{3}}{2974571600\pi}+\frac{2170050078297680341903318624256}{55495046100428560404219065625\pi^{2}}$ & $-3.6218298403...\!\times\!10^{-4}$ & \tabularnewline
\hline 
30 & $2-\frac{467243458641\sqrt{3}}{43131288200\pi}+\frac{542623285080380647450979017997312}{13840464497446882964812234966875\pi^{2}}$ & $-2.1956839098...\!\times\!10^{-4}$ & \tabularnewline
32 & $2-\frac{58010005151019\sqrt{3}}{5348279736800\pi}+\frac{628211606000412550149462722832498688}{15993372612894568794269751930170625\pi^{2}}$ & $-1.3245332534...\!\times\!10^{-4}$ & \tabularnewline
34 & $2-\frac{2002146040431\sqrt{3}}{184423439200\pi}+\frac{459908194551111759301815187699597312}{11692633759006953656314860654830625\pi^{2}}$ & $-7.9584741003...\!\times\!10^{-5}$ & \tabularnewline
36 & $2-\frac{3057909919677\sqrt{3}}{281488407200\pi}+\frac{79704452280575700446509958198372859904}{2024384658142737243029979541373008875\pi^{2}}$ & $-4.7664074481...\!\times\!10^{-5}$ & \tabularnewline
38 & $2-\frac{2150738112669759\sqrt{3}}{197886350261600\pi}+\frac{135910299223745283477388289143412323254272}{3449444910914269381215557244839534859375\pi^{2}}$ & $-2.8470483374...\!\times\!10^{-5}$ & \tabularnewline
\hline 
40 & $2-\frac{2151488933577441\sqrt{3}}{197886350261600\pi}+\frac{46533008661103403816377352659651893614608384}{1180400048514862982251963689184088828878125\pi^{2}}$ & $-1.6968201775...\!\times\!10^{-5}$ & \tabularnewline
42 & $2-\frac{88233570903905541\sqrt{3}}{8113340360725600\pi}+\frac{515591346932810387058192148814293728901860687872}{13073942308772263125282428112304512998821415625\pi^{2}}$ & $-1.0094148907...\!\times\!10^{-5}$ & \tabularnewline
44 & $2-\frac{3794753074625697753\sqrt{3}}{348873635511200800\pi}+\frac{288296736293320801496880337095858185765666520301568}{7308333750603695087032877314778222766341171334375\pi^{2}}$ & $-5.9954566098...\!\times\!10^{-6}$ & \tabularnewline
46 & $2-\frac{3795273393514721379\sqrt{3}}{348873635511200800\pi}+\frac{2519982640616868219914608548886634871739938807218176}{63868481907449683151896014794366207653677192965625\pi^{2}}$ & $-3.5562826544...\!\times\!10^{-6}$ & \tabularnewline
48 & $2-\frac{17839580049686321991\sqrt{3}}{1639706086902643760\pi}+\frac{42052273733412765929226898036759676113761323328682852352}{1065645620625797963389385006844000174701603964631453125\pi^{2}}$ & $-2.1070583282...\!\times\!10^{-6}$ & \tabularnewline
\hline 
50 & $2-\frac{624431461457604651363\sqrt{3}}{57389713041592531600\pi}+\frac{6361365531738335775999842639225946934718354708464415014912}{161185293993375696750424818595196090424665809274295073875\pi^{2}}$ & $-1.2471932532...\!\times\!10^{-6}$ & \tabularnewline
\hline 
\end{tabular}
\par\end{centering}
\caption{Fourth cluster integral, $B_{4}$, for different integer dimensions
up to $d=50$. In the third column it is shown the eleven figures
decimal expansion with the corresponding bibliographic reference where
the exact value was calculated for the first time.\label{tab:B4}}
\end{table}
 In Table \ref{tab:DecimValues.} we show some of the exactly evaluated
$B_{4}/B_{2}^{3}$ for even dimension $50<d\leq1000$. Note that the
length of the expression for the exact value increases when increasing
$d$. Therefore, it becomes so long that is inconvenient to write
it here. For this reason, we only show the decimal expansion of the
exact value truncated at eleven figures for $d=60,$ $70,$ $80,$
$90,$ $100,$ $200,$ $300,$ $500,$ and $1000$. We also include
the decimal expansion for some of the $d$ values presented in Table
\ref{tab:B4}. For comparison, the MonteCarlo numerical calculation
of Clisby and McCoy\citep{Clisby_2004b} and the high precision results
of Zhang\citep{Zhang_2014} (that reaches a maximum dimension of $d=100$),
are also presented. We observe, on one hand that those $B_{4}$ values
from Ref. \citep{Clisby_2004b} coincide with the exact result. On
the other hand, even when several $B_{4}$ values from Ref. \citep{Zhang_2014}
coincide with the exact ones, some differences appear for $d=16,18,20$.
Finally, the exact expression allows to evaluate $B_{4}$ for $d>100$
where numerical results have not been reported.
\begin{table}[h]
\centering{}%
\begin{tabular}{|c|c|c|c||c|c|c|}
\hline 
d & Exact (decimal expansion) & MC \citep{Zhang_2014} & MC \citep{Clisby_2004b} & d & Exact (decimal expansion) & MC \citep{Zhang_2014}\tabularnewline
\hline 
$14$ & $-8.2212455146...\!\times\!10^{-3}$ & $-8.22123(13)\!\times\!10^{-3}$ & $-8.220(2)\!\times\!10^{-3}$ & $70$ & $-6.3795820774...\!\times\!10^{-9}$ & $-6.37956(7)\!\times\!10^{-9}$\tabularnewline
\hline 
$16$ & $-5.7523349690...\!\times\!10^{-3}$ & $-5.75247(6)\!\times\!10^{-3}$ & $-$ & $80$ & $-4.5155630086...\!\times\!10^{-10}$ & $-4.51561(6)\!\times\!10^{-10}$\tabularnewline
\hline 
$18$ & $-3.8229418415...\!\times\!10^{-3}$ & $-3.82300(3)\!\times\!10^{-3}$ & $-$ & $90$ & $-3.1890775288...\!\times\!10^{-11}$ & $-3.18908(3)\!\times\!10^{-11}$\tabularnewline
\hline 
$20$ & $-2.4621254986...\!\times\!10^{-3}$ & $-2.46217(3)\!\times\!10^{-3}$ & $-2.4621(7)\!\times\!10^{-3}$ & $100$ & $-2.2508216158...\!\times\!10^{-12}$ & $-2.25081(2)\!\times\!10^{-12}$\tabularnewline
\hline 
$30$ & $-2.1956839098...\!\times\!10^{-4}$ & $-2.19569(2)\!\times\!10^{-4}$ & $-2.196(1)\!\times\!10^{-4}$ & $200$ & $-7.2741818933...\!\times\!10^{-24}$ & $-$\tabularnewline
\hline 
$40$ & $-1.6968201775...\!\times\!10^{-5}$ & $-1.69682(1)\!\times\!10^{-5}$ & $-1.697(1)\!\times\!10^{-5}$ & $300$ & $-2.5946808650...\!\times\!10^{-35}$ & $-$\tabularnewline
\hline 
$50$ & $-1.2471932532...\!\times\!10^{-6}$ & $-1.24719(1)\!\times\!10^{-6}$ & $-1.247(1)\!\times\!10^{-6}$ & $500$ & $-3.8031667590...\!\times\!10^{-58}$ & $-$\tabularnewline
\hline 
$60$ & $-8.9672516853...\!\times\!10^{-8}$ & $-8.96719(9)\!\times\!10^{-8}$ & $-$ & $1000$ & $-4.1655577159...\!\times\!10^{-115}$ & $-$\tabularnewline
\hline 
\end{tabular}\caption{Values of $B_{4}/B_{2}^{3}$ for even dimension $d>12$. Comparison
between the new exact result (conveniently truncated at eleven figures,
see the text) and those obtained through Monte Carlo numerical integration.
\label{tab:DecimValues.}}
\end{table}

It is known that $B_{4}/B_{2}^{3}$ is positive for $d\leq7$, negative
for $d\geq8$, have a minimum near $d=11$ (see Tab. \ref{tab:B4})
and goes to zero when $d$ becomes arbitrary large. The value of $d$
where $B_{4}=0$ was estimated by Luban and Baram several years ago\citep{Luban_1982}
as $d_{0}\approx7.8$ and more recently as $d_{0}=7.7320(4)$ by Clisby
and McCoy,\citep{Clisby_2004b} who also found the minimum of $B_{4}/B_{2}^{3}$
at $d_{1}=10.7583(2)$. Using the new expression for $B_{4}$ valid
for arbitrary dimension (complete star graph taken from Eq. \eqref{eq:B4d_fnal})
and standard find root and find minimum routines we obtain $d_{0}=7.6558249(1)$
and $d_{1}=10.67131326(3)$. Finally, to analyze the asymptotic behavior
of $B_{4}/B_{2}^{3}$ for large $d$ we fitted the expression 
\begin{equation}
\frac{B_{4}}{B_{2}^{3}}\approx\frac{4k_{1}}{\sqrt{\frac{\pi d}{2}}}\left(\frac{4\sqrt{3}}{9}\right)^{d+1}\:,\label{eq:asympt}
\end{equation}
for even dimensions $100\leq d\leq200$, to obtain $k_{1}=1.683$.
Indeed, a similar expression explains the asymptotic behavior of $B_{3}/B_{2}^{2}$
(see Ref. \citep{Santos2016}). 
\begin{figure}
\begin{centering}
\includegraphics[width=0.5\columnwidth]{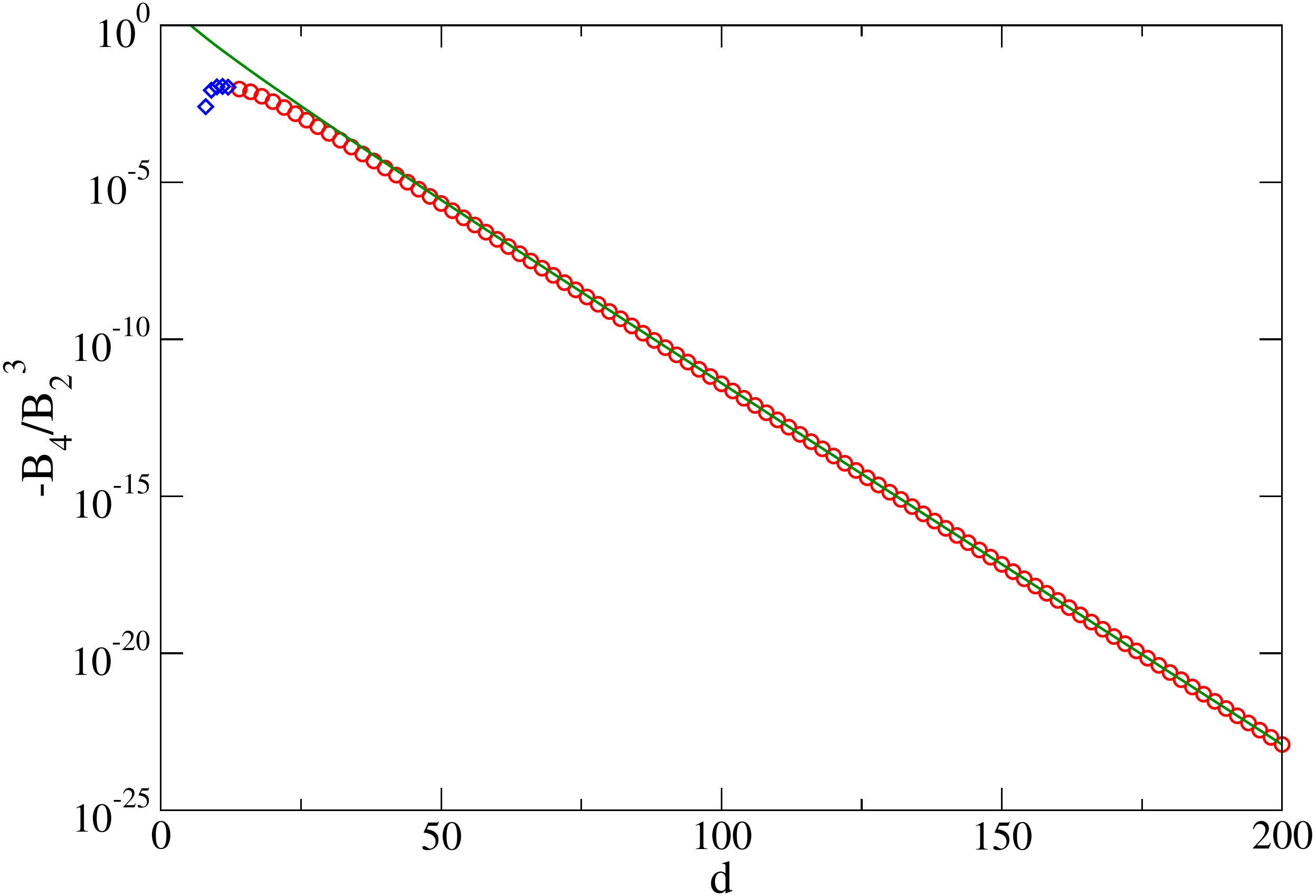}
\par\end{centering}
\caption{Asymptotic behavior of $B_{4}/B_{2}^{3}$ with dimension. Symbols
draw the exact values for integer $d$. \label{fig:B4_asintota}}
\end{figure}
In Fig. \eqref{fig:B4_asintota} it is shown the asymptotic behavior
of $B_{4}/B_{2}^{3}$ with the dimension, symbols draw the exact values
for integer dimensions where $B_{4}/B_{2}^{3}<0$. In blue diamonds
we plot the known values for $d=8$, $9$, $10$, $11$ and $12$
while in red circles we plot the new results for even dimensions from
$d=14$ to $d=200$.

\section{Final remarks\label{sec:Final-remarks}}

We return to the properties of $B_{3}$ to gain some feedback on the
$B_{4}$ form given in Eq. \eqref{eq:FrB3}. It is known that $B_{3}(d)>0$
and $\underset{d\rightarrow\infty}{\lim}B_{3}(d)=0$. For even $d=2\nu$,
Eq. \eqref{eq:B2yB3Even} can be written as
\begin{equation}
\frac{B_{3}}{B_{2}^{2}}=\frac{4}{3}-\frac{\sqrt{3}}{\pi}a_{3}\:,\label{eq:FrB3}
\end{equation}
where $a_{3}=\sum_{k=1}^{\nu}\frac{\left(\frac{3}{2}\right)^{k-1}(k-1)!}{(2k-1)\text{!!}}$
is a positive rational and $\underset{\nu\rightarrow\infty}{\lim}\frac{\sqrt{3}}{\pi}a_{3}(\nu)=\frac{4}{3}$
(see Tab. I in Ref. \citep{Clisby_2004} for the values of $B_{3}$
for positive integer dimensions $d\leq12$). It interesting to note
that $a_{4,1}=\frac{9}{2}a_{3}$. We extrapolate Eq. \eqref{eq:FrB3}
for $B_{3}$ and Eq. \eqref{eq:FrB4} for $B_{4}$ in the following
conjecture:

\paragraph{Conjecture 1:}

For any even dimension, the fifth virial coefficient of hard spheres
takes the form 
\begin{equation}
\frac{B_{5}}{B_{2}^{4}}=a_{5,0}+\frac{\sqrt{3}}{\pi}a_{5,1}+\frac{1}{\pi^{2}}a_{5,2}+\frac{\sqrt{3}}{\pi^{3}}a_{5,3}\:,\label{eq:Conj.1}
\end{equation}
with $a_{5,i}$ rational numbers.

In this work we introduced a transformation that, applied to the fourth
vertex complete star diagram, transforms three of its $f$-bonds to
Dirac delta bonds ($\delta$-bonds). The procedure allow us to obtain
a new expression for the complete star diagram of $B_{4}$ for HS,
but could also be applied to other diagrams with vertex connected
by $f$-bonds and similar step functions. Additionally, we analyzed
the obtained expression and using extensively the Table of Integrals
of Gradshteyn and Ryzhik \citep{GradshteynRyzhik2007} for integrands
involving integer powers of Sine and Cosine functions, we found an
exact expression for the complete star. Finally, to obtain the exact
$B_{4}(d)$ for every even $d\geq4$, we complemented this result
with expressions for the other two Mayer graphs. Further significant
progress in HS virial expansion coefficients, besides to study $B_{4}$
in odd dimensions, should necessary concentrate in the exact evaluation
of $B_{5}$ in dimensions $d=2$ and $d=3$. 


\section{Supplementary Material\label{sec:SupplMat}}

The exact expression of $B_{4}(d)$ for $d=100$, $200$, $300$,
$500$ and $1000$ is presented in the SupplementaryMaterial\_01.txt
file. Also, I include a computer algebra program that allows to evaluate
$B_{4}(d)$ for even dimensions in SupplementaryMaterial\_02.

\appendix

\section{Explicit form of \protect\includegraphics[width=0.5cm]{graph-b4-sqd}\label{Appsec:B4sqrdsh}}

To evaluate this Mayer diagram we transform Eq. \eqref{eq:B4Sqrd}
as
\[
\frac{\includegraphics[width=0.5cm]{graph-b4-sqd.pdf}}{B_{2}^{3}}=-2^{2\nu+6}\nu\left(\frac{((2\nu)\text{!!})^{2}}{\pi(2\nu)!}\right)^{2}\int_{0}^{\frac{1}{2}}x^{2\nu-1}H^{2}dx\:,
\]
with  $H=\intop_{0}^{\arccos x}(\sin\gamma)^{2\nu}d\gamma$, and
also
\[
H=\frac{(2\nu-1)!!}{2^{\nu}\nu!}\arccos x-\frac{x}{2\nu\sqrt{1-x^{2}}}\left[\left(1-x^{2}\right)^{\nu}+\sum_{k=1}^{\nu-1}\frac{\prod_{i=1}^{k}(2\nu-2i+1)}{2^{k}\prod_{i=1}^{k}(\nu-i)}\left(1-x^{2}\right)^{\nu-k}\right]\:,
\]
where we used Eq. \eqref{eq:IntSinpar}. We integrate by parts to
obtain
\begin{equation}
\int_{0}^{\frac{1}{2}}x^{2\nu-1}H^{2}dx=\left.\frac{x^{2\nu}}{2\nu}H^{2}\right|_{0}^{\frac{1}{2}}+\int_{0}^{\frac{1}{2}}\frac{x^{2\nu}}{\nu}\left(1-x^{2}\right)^{\nu-\frac{1}{2}}H\,dx\:,\label{eq:ap4}
\end{equation}
where the first term on the right gives $\frac{2^{-2\nu}}{2\nu}\left\{ \frac{(2\nu-1)!!}{2^{\nu}\nu!}\frac{\pi}{3}-\frac{\sqrt{3}}{6\nu}\left[\left(\frac{3}{4}\right)^{\nu}+\sum_{k=1}^{\nu-1}\frac{\prod_{i=1}^{k}(2\nu-2i+1)}{2^{k}\prod_{i=1}^{k}(\nu-i)}\left(\frac{3}{4}\right)^{\nu-k}\right]\right\} ^{2}$.
To solve the remaining integral on the right of Eq. \eqref{eq:ap4}
we change variable by replacing $x=\cos y$ to obtain 
\begin{eqnarray*}
 &  & \intop_{0}^{\frac{1}{2}}\frac{x^{2\nu}}{\nu}\left(1-x^{2}\right)^{\nu-\frac{1}{2}}H\,dx=\\
 &  & =\frac{1}{\nu}\intop_{\frac{\pi}{3}}^{\frac{\pi}{2}}\left(\sin y\cos y\right)^{2\nu}\left\{ \frac{(2\nu-1)!!}{2^{\nu}\nu!}y-\frac{\cos y}{2\nu\sin y}\left[\left(\sin y\right)^{2\nu}+\sum_{k=1}^{\nu-1}\frac{\prod_{i=1}^{k}(2\nu-2i+1)}{2^{k}\prod_{i=1}^{k}(\nu-i)}\left(\sin y\right)^{2(\nu-k)}\right]\right\} dy\:.
\end{eqnarray*}
Here, there are terms that can be written using $D\left(2l+1,2k+1\right)$
analyzed in Eq. \eqref{eq:DimparX2}, and a term $\intop_{\frac{\pi}{3}}^{\frac{\pi}{2}}y\left(\sin y\cos y\right)^{2\nu}dy$
$=2^{-2\nu-2}\intop_{\frac{2\pi}{3}}^{\pi}y\left(\sin y\right)^{2\nu}dy$.
This last integral is solved using Eq. \eqref{eq:xlSnxn} and gives
\begin{eqnarray*}
\intop_{\frac{2\pi}{3}}^{\pi}y\left(\sin y\right)^{2\nu}dy & = & \frac{5\pi^{2}}{18}\frac{(2\nu-1)\text{!!}}{(2\nu)!!}-\frac{\pi\sqrt{3}}{9}\left[\frac{\left(\frac{3}{4}\right)^{\nu}}{\nu}+\sum_{k=1}^{\nu-1}\frac{\left(\frac{3}{4}\right)^{\nu-k}\prod_{i=0}^{k-1}(2\nu-2i-1)}{2^{k}(\nu-k)\prod_{i=0}^{k-1}(\nu-i)}\right]\\
 &  & -\frac{1}{4}\left[\frac{\left(\frac{3}{4}\right)^{\nu}}{\nu^{2}}+\sum_{k=1}^{\nu-1}\frac{\left(\frac{3}{4}\right)^{\nu-k}\prod_{i=0}^{k-1}(2\nu-2i-1)}{2^{k}(\nu-k)^{2}\prod_{i=0}^{k-1}(\nu-i)}\right]\:.
\end{eqnarray*}
 Thus, one finds
\begin{eqnarray*}
\intop_{0}^{\frac{1}{2}}x^{2\nu-1}H^{2}dx & = & \frac{2^{-2\nu}}{2\nu}\biggl\{\frac{(2\nu-1)!!}{(2\nu)!!}\frac{\pi}{3}-\frac{\sqrt{3}}{6\nu}\biggl[\left(\frac{3}{4}\right)^{\nu}+\sum_{k=1}^{\nu-1}\frac{\prod_{i=1}^{k}(2\nu-2i+1)}{2^{k}\prod_{i=1}^{k}(\nu-i)}\left(\frac{3}{4}\right)^{\nu-k}\biggr]\biggr\}^{2}\\
 &  & +\frac{(2\nu-1)!!}{2^{2(\nu+1)}\nu(2\nu)!!}\biggl\{\frac{5\pi^{2}(2\nu-1)\text{!!}}{18(2\nu)!!}-\frac{\pi\sqrt{3}}{9}\biggl[\frac{\left(\frac{3}{4}\right)^{\nu}}{\nu}+\sum_{k=1}^{\nu-1}\frac{\left(\frac{3}{4}\right)^{\nu-k}\prod_{i=0}^{k-1}(2\nu-2i-1)}{2^{k}(\nu-k)\prod_{i=0}^{k-1}(\nu-i)}\biggr]\\
 &  & -\frac{1}{4}\biggl[\frac{\left(\frac{3}{4}\right)^{\nu}}{\nu^{2}}+\sum_{k=1}^{\nu-1}\frac{\left(\frac{3}{4}\right)^{\nu-k}\prod_{i=0}^{k-1}(2\nu-2i-1)}{2^{k}(\nu-k)^{2}\prod_{i=0}^{k-1}(\nu-i)}\biggr]\biggr\}\\
 &  & +\frac{1}{2\nu^{2}}\biggl\{\frac{(2\nu-1)!\,l!}{2((2\nu-1)+l+1)!}+\sum_{k=1}^{\nu-1}\frac{\prod_{i=1}^{k}(2(2\nu-1)-2i+1)}{2^{k}\prod_{i=1}^{k}((2\nu-i)}\\
 &  & +D\left(2\nu+1,4\nu-1\right)+\sum_{k=1}^{\nu-1}\frac{\prod_{i=1}^{k}(2\nu-2i+1)}{2^{k}\prod_{i=1}^{k}(\nu-i)}D\left(2\nu+1,4\nu-2k-1\right)\biggr\}\:.
\end{eqnarray*}
The final result is given in Eq. \eqref{eq:B4SqrdE}.

\section{Origin of $\cos\varphi$ term\label{Appsec:CosTerm}}

Here we derive the identity Eq. \eqref{eq:cosIdent} showing the origin
of the $\cos\varphi$ term. Let us start from Eqs. (\ref{eq:B4s-Der1a},
\ref{eq:B4s-Der1b})

\[
\includegraphics[width=0.5cm]{graph-b4-sgm.pdf}=\sigma^{d-1}{}_{*}\includegraphics[width=0.5cm]{graph-b4-sgmI.pdf}\:,
\]
we apply the derivative $\frac{\partial}{\partial\sigma}$ to both
sides, to obtain
\begin{equation}
-\includegraphics[width=0.5cm]{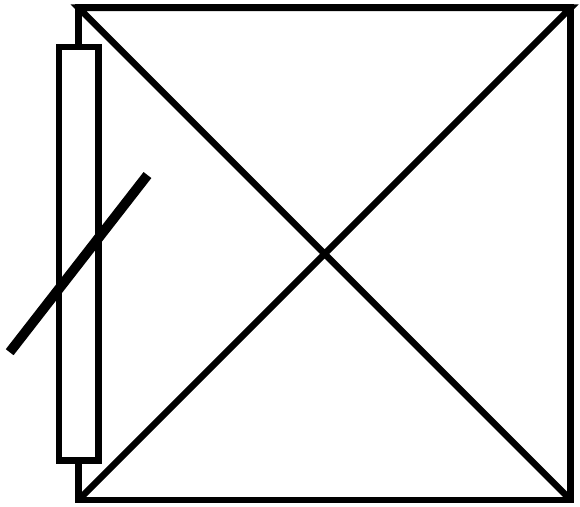}-4\:\includegraphics[width=0.5cm]{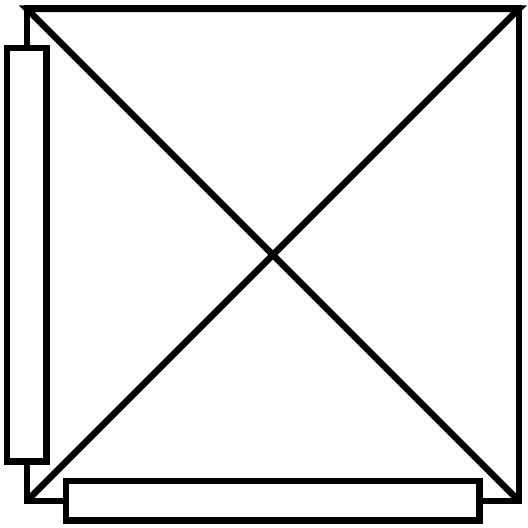}-\includegraphics[width=0.5cm]{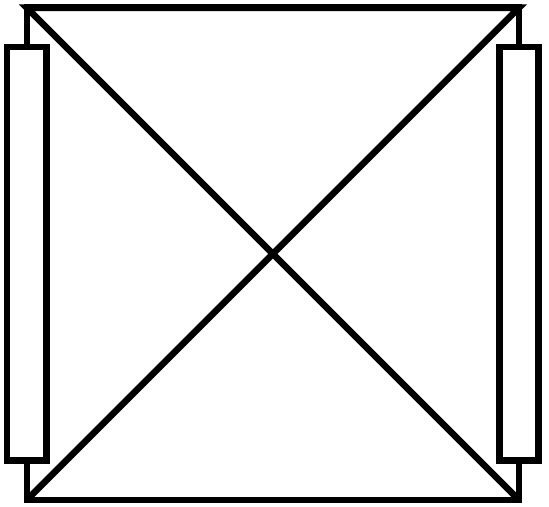}=(d-1)\sigma^{d-2}{}_{*}\includegraphics[width=0.5cm]{graph-b4-sgmI.pdf}-\sigma^{d-1}\left(2\:{}_{*}\includegraphics[width=0.5cm]{graph-b4-sgm1I1c.pdf}+2\:{}_{*}\includegraphics[width=0.5cm]{graph-b4-sgm1I1cA.pdf}+{}_{*}\includegraphics[width=0.5cm]{graph-b4-sgm1aI1c.pdf}\right)\:.\label{eqAp:cos02}
\end{equation}
Here the dashed bond corresponds to $\delta'(\sigma-r)$, being $f''=\frac{\partial^{2}f}{\partial^{2}r}=-\delta'(\sigma-r)$.
Some terms on the left are readily integrated on, $\includegraphics[width=0.5cm]{graph-b4-sgm2.pdf}=\sigma^{d-1}{}_{*}\includegraphics[width=0.5cm]{graph-b4-sgm1I1c.pdf}$
and $\includegraphics[width=0.5cm]{graph-b4-sgm2a.pdf}=\sigma^{d-1}{}_{*}\includegraphics[width=0.5cm]{graph-b4-sgm1aI1c.pdf}$.
For the diagram with a dashed bond we have
\begin{eqnarray}
\includegraphics[width=0.5cm]{graph-b4-sgmDsh.pdf} & = & -\int f'_{12}f_{13}f_{14}f_{23}f_{24}f_{34}(d-1)r_{12}^{d-2}dr_{12}d\varOmega_{12}d\mathbf{r}_{13}d\mathbf{r}_{14}-2\int f'_{12}f_{13}f_{14}\frac{\partial f_{23}}{\partial r_{12}}f_{24}f_{34}r_{12}^{d-1}dr_{12}d\varOmega_{12}d\mathbf{r}_{13}d\mathbf{r}_{14}\nonumber \\
 & = & -\frac{d-1}{V}\int f'_{12}f_{13}f_{14}f_{23}f_{24}f_{34}\frac{1}{r_{12}}d\mathbf{r}_{1,2,3,4}-\frac{2}{V}\int f'_{12}f_{13}f_{14}\cos\varphi f'_{23}f_{24}f_{34}d\mathbf{r}_{1,2,3,4}\nonumber \\
 & = & -(d-1)\sigma^{d-2}{}_{*}\includegraphics[width=0.5cm]{graph-b4-sgmI.pdf}-2\sigma^{d-1}{}_{*}\includegraphics[width=0.5cm]{graph-b4-sgm1I1c.pdf}\left\langle \cos\varphi\right\rangle \:.\label{eqAp:cosdash}
\end{eqnarray}
We replace Eq. \eqref{eqAp:cosdash} in Eq. \eqref{eqAp:cos02} to
obtain
\begin{equation}
\includegraphics[height=0.5cm]{graph-b4-sgm1I1ctps.pdf}={}_{*}\includegraphics[width=0.5cm]{graph-b4-sgm1I1c.pdf}\left\langle 1-\cos\varphi\right\rangle \:.\label{eqAp:cosend}
\end{equation}

\section{Integration of the different terms in L\label{Appsec:IntegL}}

Here we solve the split terms of $L$ (see Eq. \eqref{eq:intY02}),
the four integrals: $\left(\sqrt{3}K+\frac{4\pi}{3}\right)\int_{0}^{\frac{\pi}{3}}\sin\left(\varphi\right)^{2m+2}d\varphi$,\linebreak{}
 $-\left(\frac{\sqrt{3}}{2}K+\frac{2\pi}{3}\right)\int_{0}^{\frac{\pi}{3}}\cos\varphi\left(\sin\varphi\right)^{2m+2}d\varphi$,
$-2\int_{0}^{\frac{\pi}{3}}\varphi\left(\sin\varphi\right)^{2m+2}d\varphi$,
and $\int_{0}^{\frac{\pi}{3}}\varphi\cos\varphi\left(\sin\varphi\right)^{2m+2}d\varphi$.
To solve the first one we use \footnote{Ref. \citep{GradshteynRyzhik2007}, Sec. 2.511 Eq.(2), at p.152.}
\begin{equation}
\int_{0}^{u}\left(\sin\varphi\right)^{2l}d\varphi=\frac{(2l-1)!!}{2^{l}l!}u-\frac{\cos u}{2l}\biggl[\left(\sin u\right)^{2l-1}+\sum_{k=1}^{l-1}\frac{\prod_{i=1}^{k}(2l-2i+1)}{2^{k}\prod_{i=1}^{k}(l-i)}\left(\sin u\right)^{2l-2k-1}\biggr]\:,\label{eq:IntSinpar}
\end{equation}
to obtain
\[
\int_{0}^{\frac{\pi}{3}}\left(\sin\varphi\right)^{2m+2}d\varphi=\frac{\frac{\pi}{3}(2m+1)!!}{2^{m+1}(m+1)!}-\frac{\sqrt{3}}{8(m+1)}\biggl[\left(\frac{3}{4}\right)^{m}+\sum_{k=1}^{m}\frac{\prod_{i=0}^{k-1}(2m-2i+1)}{2^{k}\prod_{i=0}^{k-1}(m-i)}\left(\frac{3}{4}\right)^{m-k}\biggr]\:.
\]
The second one is straightforward, it gives
\[
\int_{0}^{\frac{\pi}{3}}\cos\varphi\left(\sin\varphi\right)^{2m+2}d\varphi=\frac{\sqrt{3}}{4m+6}\left(\frac{3}{4}\right)^{m+1}\:.
\]
To solve the third one, $-2\int_{0}^{\frac{\pi}{3}}\varphi\left(\sin\varphi\right)^{2m+2}d\varphi$,
we apply \footnote{Ref. \citep{GradshteynRyzhik2007}, Sec. 2.631 Eq.(2), at p.214.}
\begin{equation}
\int_{0}^{u}x^{l}\left(\sin x\right)^{n}dx=\frac{u^{l-1}\left(\sin u\right)^{n-1}}{n^{2}}\left(l\sin u-n\,u\cos u\right)+\frac{n-1}{n}\int_{0}^{u}x^{l}\left(\sin x\right)^{n-2}dx-\frac{l(l-1)}{n^{2}}\int_{0}^{u}x^{l-2}\left(\sin x\right)^{n}dx\,,\label{eq:xlSnxn}
\end{equation}
iteratively, to obtain
\begin{eqnarray}
\int_{0}^{u}x\left(\sin x\right)^{2n}dx & = & \frac{u^{2}(2n-1)\text{!!}}{2\left(2^{n}n!\right)}+\frac{1}{4}\biggl[\frac{(\sin u)^{2n}}{n^{2}}+\sum_{k=1}^{n-1}\frac{\prod_{i=0}^{k-1}(2n-2i-1)}{2^{k}(n-k)^{2}\prod_{i=0}^{k-1}(n-i)}(\sin u)^{2(n-k)}\biggr]\nonumber \\
 &  & -\frac{u\cos u}{2\sin u}\biggl[\frac{(\sin u)^{2n}}{n}+\sum_{k=1}^{n-1}\frac{\prod_{i=0}^{k-1}(2n-2i-1)}{2^{k}(n-k)\prod_{i=0}^{k-1}(n-i)}(\sin u)^{2(n-k)}\biggr]\:,\label{eq:xSnx2n}
\end{eqnarray}
and finally
\begin{eqnarray*}
\int_{0}^{\frac{\pi}{3}}\varphi\left(\sin\varphi\right)^{2m+2}d\varphi & = & \frac{\pi^{2}(2m+1)\text{!!}}{18(2m+2)!!}+\frac{1}{4}\biggl[\frac{\left(\frac{3}{4}\right)^{m+1}}{(m+1)^{2}}+\sum_{k=1}^{m}\frac{\left(\frac{3}{4}\right)^{m-k+1}\prod_{i=0}^{k-1}(2m-2i+1)}{2^{k}(m-k+1)^{2}\prod_{i=0}^{k-1}(m-i+1)}\biggr]\\
 &  & -\frac{\pi\sqrt{3}}{18}\biggl[\frac{\left(\frac{3}{4}\right)^{m+1}}{m+1}+\sum_{k=1}^{m}\frac{\left(\frac{3}{4}\right)^{m-k+1}\prod_{i=0}^{k-1}(2m-2i+1)}{2^{k}(m-k+1)\prod_{i=0}^{k-1}(m-i+1)}\biggr]\:.
\end{eqnarray*}
To solve the fourth one we use \footnote{Ref. \citep{GradshteynRyzhik2007}, Sec. 2.631 Eq.(1), at p.214.}
\begin{eqnarray*}
\int_{0}^{u}x\left(\sin x\right)^{p}\left(\cos x\right)^{q}\,dx & = & \frac{1}{\left(p+q\right)^{2}}\left[\left(p+q\right)u\left(\sin u\right)^{p+1}\left(\cos x\right)^{q-1}+\left(\sin u\right)^{p}\left(\cos x\right)^{q}\right.\,,\\
 &  & \left.-p\int_{0}^{u}\left(\sin x\right)^{p-1}\left(\cos x\right)^{q-1}dx+\left(q-1\right)\left(p+q\right)\int_{0}^{u}\left(\sin x\right)^{p}\left(\cos x\right)^{q-2}dx\right]\:.
\end{eqnarray*}
In particular, for $q=1$ it reduces to
\[
\int_{0}^{u}x\left(\sin x\right)^{p}\cos x\,dx=\frac{1}{(p+1)^{2}}\left[(p+1)u\left(\sin u\right)^{p+1}+\left(\sin u\right)^{p}\cos u-p\int_{0}^{u}\left(\sin x\right)^{p-1}dx\right]\,,
\]
which gives the result
\[
\int_{0}^{\frac{\pi}{3}}\varphi\cos\varphi\left(\sin\varphi\right)^{2m+2}d\varphi=\frac{\pi\sqrt{3}\left(\frac{3}{4}\right)^{m+1}}{6(2m+3)}+\frac{\left(\frac{3}{4}\right)^{m+1}}{2(2m+3)^{2}}-\frac{2(m+1)}{(2m+3)^{2}}\int_{0}^{\pi/3}\left(\sin x\right)^{2m+1}dx\:.
\]
To solve the last integral we used Eq. \eqref{eq:intSnxiCsxq} (with
$q=0$) to obtain 
\[
\int_{0}^{\frac{\pi}{3}}\left(\sin x\right)^{2n+1}dx=\frac{2^{n}n!}{\left(2n+1\right)!!}-\frac{1}{2(2n+1)}\biggl[\left(\frac{3}{4}\right)^{n}+\sum_{k=1}^{n}\frac{2^{k}\prod_{i=0}^{k-1}(n-i)}{\prod_{i=0}^{k-1}(2n-2i-1)}\left(\frac{3}{4}\right)^{n-k}\biggr]\,.
\]
\bibliographystyle{unsrt}

\end{document}